\documentclass[%
 reprint,
superscriptaddress,
nofootinbib,
 amsmath,amssymb,
 aps,
pra,
floatfix,
]{revtex4-2}

\usepackage{graphicx}
\usepackage{dcolumn}
\usepackage{bm}

\usepackage{xcolor}
\usepackage{slashed}
\usepackage{braket}
\usepackage{comment}

\newcommand{\w}{\omega}
\newcommand{\W}{\Omega}

\newcommand{\bb}{\langle}
\newcommand{\rr}{\rangle}
\newcommand{\g}{\gamma}
\newcommand{\sg}{\sigma}
\newcommand{\p}{\partial}
\newcommand{\mb}{\mathbb}
\newcommand{\al}{\alpha}
\newcommand{\bs}{\mathbf}
\newcommand{\m}{\mathcal}

\newcommand{\ep}{\varepsilon}
\newcommand{\eps}{\epsilon}

\newcommand{\n}{\nabla}
\newcommand{\bt}{\beta}

\newcommand{\Tb}{\mathbb{T}}
\newcommand{\B}{\mathbb{B}}
\newcommand{\J}{\mathbb{J}}
\newcommand{\A}{\mathbb{A}}

\newcommand{\vct}[1]{\mathbf{#1}}
\newcommand{\dgcomm}[1]{\textcolor{red}{DG: #1}}

\begin{document}

\preprint{APS/123-QED}

\title{Heat Transfer and Torque in Enclosing  Cylindrical Configurations\\ with Nonreciprocal Materials}
\author{Dhruv Shah}
\email[Correspondence email address: ]{dhruvashah2004@gmail.com}
\affiliation{Department of Physics, Massachusetts Institute of Technology, Cambridge, MA 02139, USA}
\author{Kiryl Asheichyk}
\email{kiryl.asheichyk@gmail.com}
\affiliation{Department of Theoretical Physics and Astrophysics, Belarusian State University, 5 Babruiskaya Street, 220006 Minsk, Belarus}
\author{David Gelbwaser-Klimovsky}
\email{dgelbi@technion.ac.il}
\affiliation{Schulich Faculty of Chemistry and Helen Diller Quantum Center, Technion--Israel Institute of Technology, Haifa 3200003, Israel}
\author{Noah Graham}
\email{ngraham@middlebury.edu}
\affiliation{Department of Physics, Middlebury College, Middlebury, VT 05753, USA}
\author{Mehran Kardar}
\email{kardar@mit.edu}
\affiliation{Department of Physics, Massachusetts Institute of Technology, Cambridge, MA 02139, USA}
\author{Matthias Kr\"uger}
\email{matthias.kruger@uni-goettingen.de}
\affiliation{Institute for Theoretical Physics, Georg-August-Universit\"at G\"ottingen, 37077 G\"ottingen, Germany}
\date{\today}

\begin{abstract}
Electromagnetic fluctuations can transfer not only energy but also angular momentum, leading to forces, torques, heat currents, and friction in out-of-equilibrium setups. In enclosing configurations, we show that if at least one of two objects is rotationally symmetric, the torque is bounded by heat transfer, since both arise from photon transfers with angular momentum $\hbar n$ and energy $\hbar\omega$. With only one object assumed to be rotationally symmetric, it may be possible to obtain a nonzero torque with reciprocal media, but nonreciprocal media are required to break the symmetry between $n$ and $-n$ and produce a nonzero torque if both objects are rotationally symmetric. 
We then specialize to concentric cylinders with a nonreciprocal dielectric response and use Rytov fluctuational electrodynamics to express  heat transfer and torque in terms of an angular-momentum-resolved flux density, $\Phi_n(\omega)$. 
We also analyze the conditions for stable levitation of the inner cylinder using the proximity force approximation, in the process obtaining a new analytic formula for the normal Casimir force between dilute plates at different temperatures.  Finally, to find the extracted work in a contactless engine setup, we compute the fluctuation-induced friction for a slowly rotating inner cylinder, and we find a bound between torque, friction, and heat transfer. Due to this bound, the efficiency of the heat engine remains bounded by the Carnot limit.
\end{abstract}

\maketitle


\section{Introduction}

Because electromagnetic fluctuations carry both energy and momentum, they can mediate forces, torques, heat currents, and friction. 
A well-known  example is the attractive force between perfect mirrors separated by vacuum, caused by zero-point fluctuations of the electromagnetic (EM) field, as first calculated by Casimir~\cite{cas}. Lifshitz generalized these calculations to realistic materials and finite temperatures~\cite{liftz}. For objects held at different temperatures, the corresponding nonequilibrium fluctuations are described by the Rytov formalism~\cite{rytov1959theory}, which predicts phenomena such as near-field radiative heat transfer~\cite{Polder_HT_plates}, which has been observed in a variety of experiments over the past decades~\cite{biggerheat,HT_exp2, HT_exp3,HT_exp4}.

Out of equilibrium, fluctuation-induced forces display phenomena absent at thermal equilibrium~\cite{Antezza2008}, including stable configurations~\cite{Kruger_2011} and propulsion forces~\cite{Muller_2016,reid2017,nr,gyrodparticle,Henkes2025,Jiao2026}. Such forces suggest the broader question addressed here: under what symmetry and material conditions can nonequilibrium EM fluctuations transfer angular momentum, produce a steady torque, and perform work? Previous proposals used this mechanism to design contactless heat engines made of parallel plates~\cite{nr} or nanoparticles~\cite{gyrodparticle}. On general grounds of symmetry and thermodynamic time reversibility, extracting work from a tangential force requires bodies at different temperatures in a geometry that breaks left-right symmetry. For parallel plates, optical anisotropy, such as an oblique optical axis, breaks this spatial symmetry, but electromagnetic nonreciprocity of at least one plate is still required to obtain a tangential force~\cite{nr}. This result motivates the question of whether nonreciprocity is also essential for steady fluctuation-induced torques in other geometries. More broadly, optically nonreciprocal media have been shown to display unusual thermodynamic behavior, including persistent heat currents in thermal equilibrium~\cite{zhu_persistent_2016, Henkes2025}, violations of Kirchhoff's law~\cite{zhu_near-complete_2014}, deviations from Green--Kubo relations~\cite{herz_green-kubo_2019}, the photon thermal Hall effect~\cite{ben-abdallah_photon_2016}, and giant magnetoresistance for heat flux~\cite{latella_giant_2017}.

In this work, we build on the recent manuscript \cite{shah2026}, providing various additional results and discussion. We focus on two coaxial cylinders separated by vacuum, but also investigate the conditions for a nonzero torque and its relation to heat transfer more generally. If at least one of the objects is rotationally symmetric, we find a bound between torque and heat. If both objects are rotationally symmetric, the torque vanishes for reciprocal media, so that at least one of them must be optically nonreciprocal for a nonzero torque. We analyze in detail the case of two concentric cylinders at different temperatures, for which we calculate heat transfer and torque in terms of an angular-momentum-resolved heat flux density, $\Phi_n(\omega)$, and obtain analytic results in the thin-cylinder limit.

We also investigate the conditions for stable levitation of the inner cylinder. We show, using the proximity force approximation (PFA), that any repulsive force in the geometry of parallel plates yields a stable configuration for the case of two cylinders. Such repulsion for parallel plates is indeed obtained for objects at different temperatures~\cite{Bimonte2011}. By finding an analytic formula for this repulsion and quantifying the influence of nonreciprocity, we demonstrate possibilities for a stable cylinder configuration.

Finally, when the inner cylinder rotates while the outer cylinder is fixed, we compute the resulting torque and heat transfer under rotation~\cite{linearresponse, greenkubo}, finding a bound between friction, torque and heat transfer. We obtain the efficiency of the proposed heat engine which is bounded by the Carnot efficiency.

The manuscript is organized as follows: In Sec.~\ref{sec:General}, we consider propulsive torque and heat transfer in the general case of an enclosing geometry, demonstrating a bound between the two and the need for nonreciprocity to obtain a nonzero torque. In Sec.~\ref{sec:cyl}, we consider the specific geometry of two concentric cylinders made of nonreciprocal media. We give explicit formulae for heat transfer and torque, and discuss their symmetries. In Sec.~\ref{sec:stab}, we find conditions for the configuration of the inner cylinder to be stable using the proximity force approximation. We derive a concise formula for the Casimir force between two parallel plates, demonstrating the possibility of repulsion. In Sec.~\ref{sec:rot}, we analyze the torque and heat transfer when the inner cylinder rotates, and find the bound between torque, friction and heat transfer along with the associated efficiency of the heat engine. We conclude in Sec.~\ref{sec:concl}.
A number of important technical details, such as computation of scattering matrices for nonreciprocal media in a cylindrical basis, along with corresponding heat and torque formulae, are relegated to Appendices.

\section{Heat transfer and  torque in enclosing geometries}\label{sec:General}
\begin{figure}[!htbp]
    \centering
    \includegraphics[trim=0 0 0 0, clip,height=4cm]{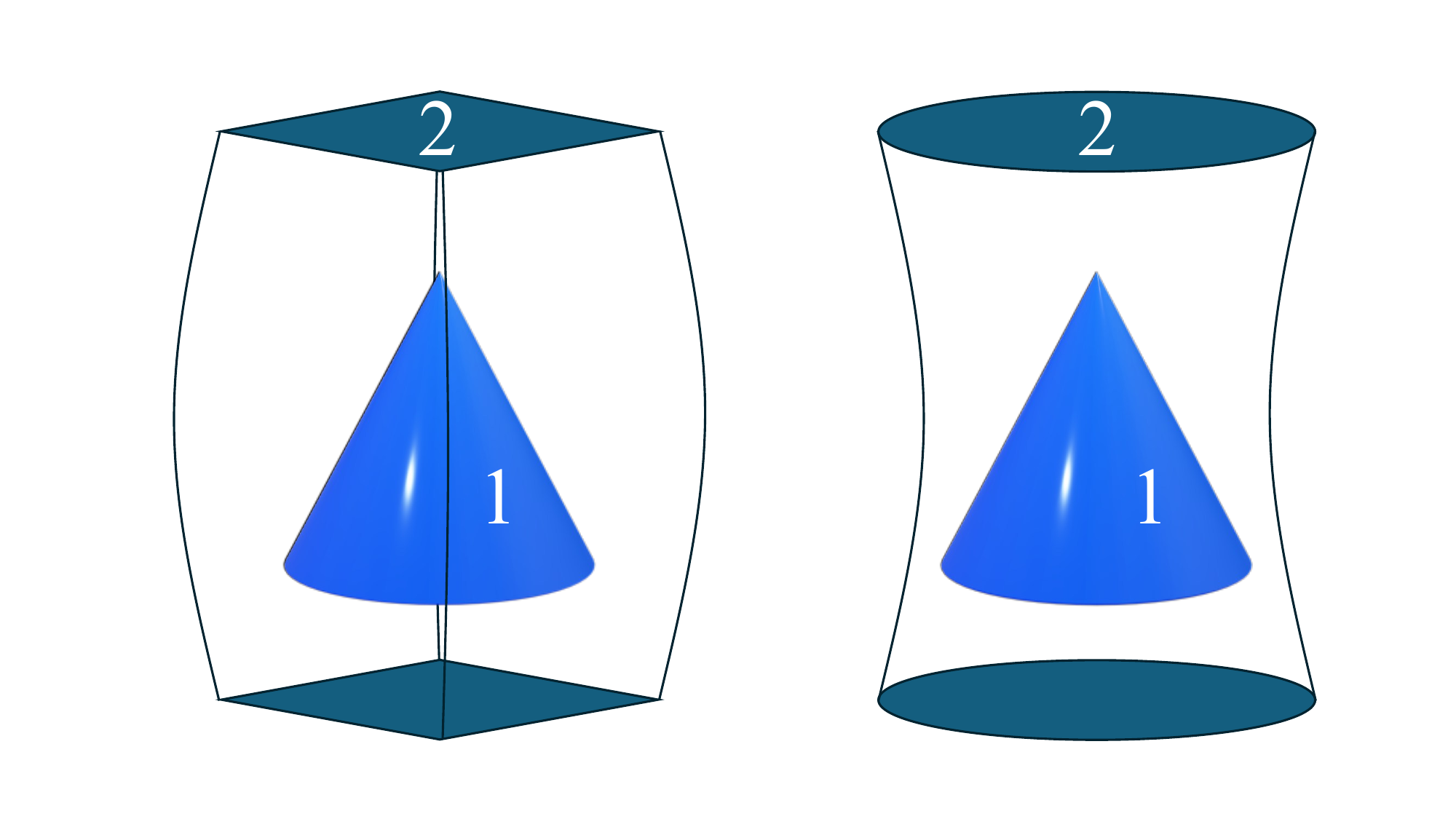}
    \caption{One object enclosing another one. 
        Left: Inner rotationally symmetric object enclosed by another arbitrary object. In this case, there is a bound between torque acting on the inner one and the radiative heat transfer between the objects, derived in Sec.~\ref{sec:bound}. Right: Inner rotationally symmetric object enclosed by another rotationally symmetric object. In this case, there is no torque for reciprocal materials, as shown in Sec.~\ref{sec:zero}.}
    \label{fig:setup0}
\end{figure}
Before analyzing the specific setup of two concentric cylinders, we consider in this section the more general case of a rotationally symmetric object (object 1) enclosed by a second one (object 2), as depicted in Fig.~\ref{fig:setup0}. The enclosing object is assumed to be thick compared to the penetration depths of electromagnetic waves, so that the electromagnetic field in the vacuum gap between the two objects is unaffected by the field surrounding the outer object.

\subsection{Fluctuating--field correlator for two objects}

The Rytov formalism considers fluctuating current sources for Maxwell's equations, quantified by the fluctuation-dissipation theorem~\cite{eckhardt}. 
The electric field $\bs E$ is then related to these fluctuations through the system's Green function $\mb G$. 
For global equilibrium at inverse temperature $T$, averaging over these fluctuations leads to the spectral density of the symmetrized correlator of electric fields as~\cite{eckhardt}
\begin{align}\label{equilibrium ee}
    \bb \bs E \otimes \bs E^* \rr_{\w}^\text{eq} &= \frac{8\pi\hbar\omega^2}{c^2} \,\text{sgn}(\w) \left[\frac{1}{2} + \frac{1}{e^{\bt \hbar|\w|}-1}\right] \nonumber \\ &\times \frac{\mb G(\w) - \mb G^\dagger(\w)}{2i},
\end{align} 
where $\bt \equiv \frac{1}{k_{\textrm{B}}T}$ and $ k_{\textrm{B}} $ is Boltzmann's constant.
We define $a_\bt(\w)\equiv \frac{8\pi\hbar}{c^2} \, \text{sgn}(\w) \left[\frac{1}{2}+\frac{1}{e^{\hbar\bt |\w|}-1} \right] \equiv \frac{4\pi\hbar}{c^2}\text{sgn}(\w) + \frac{8\pi \hbar}{c^2}\text{sgn}(\w)\nu_\bt(\w)$, where $\hbar$ is Planck's constant and $c$ is the speed of light in vacuum. The first term describes  zero-point fluctuations, while the latter is the thermal contribution with Bose-Einstein distribution $\nu_\bt(\w)$.

In Eq.~\eqref{equilibrium ee}, $\mb G$ is the dyadic Green's function, given by~\cite{Jackson} 
\begin{align}
    \mb G = \left[ \n \times \n \times (\ldots) -\frac{\w^2}{c^2} - \mb V\right]^{-1},
\end{align}
 where $\mb V$ is the material potential given by $\mb V = \frac{\w^2}{c^2}(\bs \eps-\mb I)$. 
For the system of two objects in Fig.~\ref{fig:setup0}, we split $\mb V = \mb V_1 + \mb V_2$, where $\mb V_i$ is only {non-zero} within object $i$. The total correlator in Eq.~\eqref{equilibrium ee} can be decomposed into contributions due to fluctuations in the two objects and in the environment. Because the outer enclosing object is thick compared to its skin depth, the external environment makes no contribution to the correlator in the vacuum gap between the  objects~\cite{trace, thesisgolyk}. Assuming a non-equilibrium steady state, with the two objects at different temperatures $\bt_1 \neq \bt_2$, the total correlator can be then decomposed as~\cite{trace}
\begin{align}\label{eq: 2 temp corr split}
    \bb \bs E \otimes \bs E^* \rr_\w = \bb \bs E \otimes \bs E^*\rr_{\w,1}(\bt_1) + \bb \bs E  \otimes \bs E^\dagger\rr_{\w,2}(\bt_2).
\end{align}
The correlator sourced by fluctuations in object $i$, $\bb \bs E \otimes \bs E^*\rr_{\w,i}(\bt_i)$, can be conveniently expressed in terms of its electromagnetic scattering operator $\mb T_i = \mb V_i \left(1 - \mb G_0 \mb V_i \right)^{-1}$ as~\cite{trace}
\begin{align}
    \bb \bs E \otimes \bs E^* \rr_{\w,i}&= \w^2a_{\bt_i}(\w) \mb O_{i\bar i} \mb R_i \mb O_{i\bar i}^\dagger \nonumber \\
    \mb R_i &= \mb G_0 \left[\frac{\mb T_i - \mb T_i^\dagger}{2i} - \mb T_i \text{Im} \mb G_0 \mb T_i^\dagger \right] \mb G_0^*  \, .\label{eq:rad}
\end{align} 
where $\bar i= 2$ if $i=1$ and vice versa, and $\mb O_{ij} \equiv (1+\mb G_0 \mb T_{j}) \left( 1-\mb G_0 \mb T_i \mb G_0 \mb T_{j}\right)^{-1}$ is the (extended) multiple scattering operator. 

\subsection{General trace expressions for heat  and torque}\label{sec trace torque}
For a general 2-body system with scattering operators $\mb T_1$ and $\mb T_2$, 
the heat transfer sourced by object 2 and absorbed by object 1 can be written as~\cite{trace,nr}
 \begin{align}
H_2^{(1)}
&=4\int_0^\infty \frac{d\omega}{2\pi}  \frac{\hbar\w}{e^{\hbar\omega\beta_2}-1} \,\text{Tr} \left[{\mathbb{A}}_1\mathbb{W}\mathbb{R}_2\mathbb{W}^\dagger\right],\label{eq:heat trace}
\end{align}
where $\mathbb{R}_2$ is the radiation operator defined in Eq.~\eqref{eq:rad}, $\mathbb{A}_1=\mathbb{G}_0^*\left[\frac{\mathbb{T}_1-\mathbb{T}_1^\dagger}{2i} - \mathbb{T}^\dagger_1 {\rm Im}[\mathbb{G}_0] \mathbb{T}_1\right]\mathbb{G}_0$
is the absorption operator, and $\mathbb{W}\equiv \mathbb{G}_0^{-1}(1-\mathbb{G}_0\mathbb{T}_2\mathbb{G}_0\mathbb{T}_1)^{-1}$ is the multiple scattering operator.
The torque acting on object 1 due to fluctuations in object 2 is also readily expressed in a compact notation via~\cite{Strekha22}
\begin{align}\label{eq:torque general trace}
    \tau_2^{(1)}
    &=-4\int_0^\infty \frac{d\omega}{2\pi}  
    \frac{1}{e^{\hbar\omega\beta_2}-1} \textrm{Im}\text{Tr}\left[{\mathbb{M}}_1\mathbb{W}\mathbb{R}_2\mathbb{W}^\dagger\right],
\end{align}
where 
$\mathbb{M}_1=\mathbb{G}_0^*\left[\mathbb{T}_1^\dagger\mathbb{J}_z (1+\mathbb{G}_0 \mathbb{T}_1)\right]\mathbb{G}_0$, with angular momentum $\mathbb{J}_z=\mathbb{L}_z+\mathbb{S}_z$ for orbital and spin angular momentum operators $\mathbb L_z=-i\hbar(\mathbf{r}\times\nabla)_z$ and $(\mathbb{S}_z)_{ij}=-i\hbar\varepsilon_{zij}$, respectively.

\subsection{Enclosing geometries}
For the enclosing geometries considered in Fig.~\ref{fig:setup0}, simplifications can be made to the trace expressions above. First, we consider an object with material potential $\mb V_1$ at temperature $\beta_1$, in an environment at temperature $\beta_{\text{e}}$ with dielectric response $\eps_{\text{e}}$. For $\bt_1 \neq  \bt_{\text{e}} $, the electric field correlator can be decomposed similarly to Eq.~(\ref{eq: 2 temp corr split}) into contributions from the object and the environment given by $\bb {\bf E} \otimes {\bf E}^*\rr_{\w,1} = a_{\bt_1} \,\mb G\frac{\mb V_1-\mb V_1^\dagger}{2i}\mb G^\dagger$ and $\bb {\bf E}\otimes {\bf E}^*\rr_{\w,\text{e}} = a_{\bt_{\text{e}}}\, \mb G \,\text{Im}(\mb G_0^{-1})\mb G^\dagger$, respectively~\cite{trace}. The contribution of the environment is then non-zero, because the environment covers a region of infinite volume, so that a finite value remains in the limit $\eps_{\text{e}}\rightarrow1$~\cite{shorttrace, cylradiation}.  


In the case of a finite cavity within an enclosing surrounding object (object 2), one can similarly show that the limit $\eps_\text{e} \rightarrow1$ yields a vanishing environment contribution. The radiation correlator within this cavity in absence of inner object 1 is the equilibrium correlator in Eq.~(\ref{equilibrium ee}), and the radiation operator of object 2, appearing in Eq.~\eqref{eq:heat trace}, can thus be alternatively written as
\begin{align}\label{eq: alt rad operator}
\mathbb{R}_2=\frac{\mathbb{G}_2-\mathbb{G}_2^\dagger}{2i}=\frac{\mb G_0\mathbb{T}_2\mb G_0-\mb G_0^*\mathbb{T}_2^\dagger \mb G_0^*}{2i} +  {\rm Im}\left(\mathbb{G}_0\right).
\end{align}
The absorption operator $\mathbb{A}_2$ may equivalently be found  by noting that, from conservation of energy, $H_1^{(2)}$ equals minus the self-emission of object 1, which yields
\begin{align}    \label{eq:A}
\mathbb{A}_2=\frac{\mathbb{G}_2-\mathbb{G}_2^\dagger}{2i}=\mathbb{R}_2.
\end{align}
In contrast to the general case, radiation and absorption operators are identical for an enclosing object.

For reciprocal materials, we have $\mathbb{A}_2=\mathbb{R}_2^*$~\cite{trace}, which is fulfilled by Eqs.~\eqref{eq: alt rad operator} and \eqref{eq:A} because in this case $\mathbb{A}_2= \mathbb{R}_2=\textrm{Im}[\mathbb{G}_2]$ is real. 

\subsection{A torque-heat bound}\label{sec:bound}
We assume in this section that object 1 is symmetric under rotation around the $z$-axis, i.e., $\mathbb{T}_1(\mathbf{r},\mathbf{r}')=\mathbb{R}_z\mathbb{T}_1(\mathbb{R}_z\mathbf{r},\mathbb{R}_z\mathbf{r}')\mathbb{R}_z^T$ with $\mathbb{R}_z$ the rotation matrix in three dimensions around the $z$-axis, so that $[\mathbb{J}_z,\mathbb{T}_1]=0$. For any Hermitian operator $\mb Y$, $\mbox{Im}\mbox{Tr}[\mb X \mb Y]=\frac{1}{2i}\mbox{Tr}[(\mb X-\mb X^\dagger)\mb Y]$, so we can replace $\mathbb{M}_1$ in Eq.~\eqref{eq:torque general trace} by its anti-Hermitian part. Since $\mathbb{J}_z$ is Hermitian, we can replace $\mathbb{M}_1=\mathbb{J}_z\mathbb{A}_1$, with $\mathbb{A}_1$ defined below Eq.~\eqref{eq:heat trace}, giving for the torque
\begin{align}
\tau_2^{(1)}
&=4\int_0^\infty \frac{d\omega}{2\pi}  
\frac{1}{e^{\hbar\omega\beta_2}-1} \text{Tr}\left[{\mathbb{J}}_z\mathbb{A}_1\mathbb{W}\mathbb{R}_2\mathbb{W}^\dagger\right].
\end{align}
We furthermore use that the Hermitian, positive semidefinite operator $\mathbb{A}_1$ has a unique Hermitian positive semidefinite square root, $\mathbb{B}=\sqrt{\mathbb{A}_1}$, with $\mathbb{A}_1=\mathbb{B}\mathbb{B}$. Since $\B$ commutes with $\J_z$ if $\A_1$ does, by using the cyclic property of the trace we have
\begin{align}
H_2^{(1)}
&=4\int_0^\infty \frac{d\omega}{2\pi}  \frac{\hbar\w}{e^{\hbar\omega\beta_2}-1} \,\text{Tr} \left[{\mathbb{B}}\mathbb{W}\mathbb{R}_2\mathbb{W}^\dagger\B\right], \nonumber \\
\tau_2^{(1)}
&=4\int_0^\infty \frac{d\omega}{2\pi}  
\frac{1}{e^{\hbar\omega\beta_2}-1} \text{Tr}\left[{\mathbb{J}}_z\mathbb{B}\mathbb{W}\mathbb{R}_2\mathbb{W}^\dagger\B\right]
\end{align}
for heat transfer and torque.
The Hermitian, positive semidefinite operator $\mathbb{H}\equiv\mathbb{B}\mathbb{W}\mathbb{R}_2\mathbb{W}^\dagger\B$ can be expanded in $\ket{n,\alpha}$, the eigenbasis of ${\mathbb{J}}_z$, where ${\mathbb{J}}_z\ket{n,\alpha}=\hbar n\ket{n,\alpha}$ with integer $n$, and $\alpha$ makes explicit the degeneracy of eigenvalue $\hbar n$. We obtain
\begin{align}
 h\equiv \mbox{Tr} [\mathbb{H}]
        \notag &=\sum_{n,\alpha}  \braket{n,\alpha|\mathbb{H}|n,\alpha}=\sum_{n}h_n,\\
         j\equiv\frac{1}{\hbar}\mbox{Tr} [\mathbb{J}_z\mathbb{H}]&=\sum_{n,\alpha} n \braket{n,\al|\mathbb{H}|n,\al}=\sum_{n}nh_n,
\end{align}
where we have defined 
$h_n=\sum_\alpha\braket{n,\alpha|\mathbb{H}|n,\alpha}$ with $h_n\geq0$, because $\mathbb{H}$ is positive semidefinite. The above yields a bound between torque and heat transfer, similar to what was found regarding the tangential force for translationally invariant objects~\cite{nr}. 
In practical cases, the sum over $n$ can be truncated at $n=N$. Using $|\sum_{n=-N}^N n h_n|\leq N \sum_n h_n=N\mbox{Tr}[\mathbb{H}]$, we find
\begin{align}
|j|\leq N h.
\end{align}
The typical value of $N$ is set by the radii of the two objects, and as is the bound. If the inner object is a thin cylinder, meaning its radius is small compared to all other length scales, one can restrict to $N=1$, and the bound becomes 
\begin{align}
|j|\leq h.\label{eq:bound}
\end{align}
While we leave an explicit verification for future numerical work, for concentric cylinders with radii $R_1$ and $R_2$, as shown in Fig.~\ref{fig:setup}, we may guess $N\sim R_1/(R_2-R_1)$ for $(R_2-R_1)\ll R_1$, and $N\sim \frac{\omega}{c}R_1$ for $(R_2-R_1)\gg R_1$. This would imply that torque in relation to heat transfer is largest at close proximity, where the bound diverges as $1/(R_2-R_1)$, in agreement with what was found for two parallel plates \cite{nr}.    

\subsection{Reciprocity implies zero torque}\label{sec:zero}
\begin{figure}[!htbp]
    \centering
    \includegraphics[trim=200 120 450 150, clip,width=5cm]{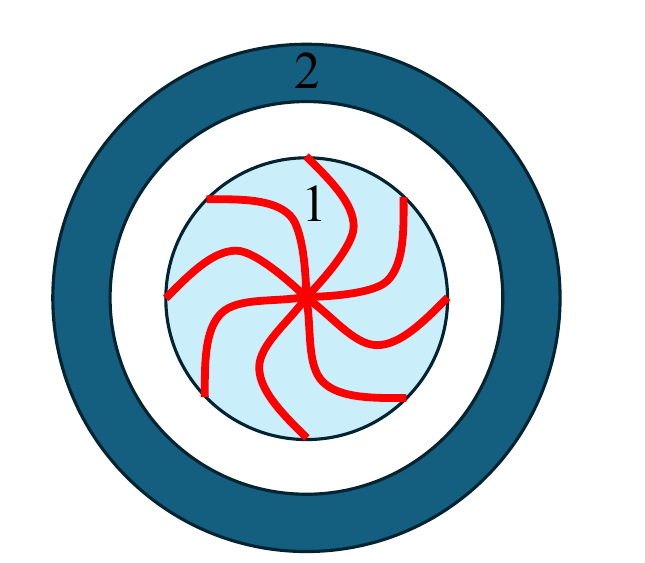}
    \caption{Concentric cylinders with rotational symmetry made from reciprocal, but possibly anisotropic material. In the illustration, the local anisotropy axes of the inner cylinder break clockwise to counter-clockwise (chiral) symmetry. While the broken chiral symmetry allows in principle a finite torque, as shown in Sec.~\ref{sec:zero}, there is no torque in this setup. }
    \label{fig:Sketch2}
\end{figure}
In this section we assume that both objects commute with $\J_z$, meaning that the $\Tb$-matrices for both objects are symmetric with respect to rotations around the $z$-axis.  
We expand $\Tb_i$ and $\mathbb{G}_0$ in regular and outgoing vector spherical waves $\ket{\vct P^{\textrm{r}}_{l,n}}$ and $\ket{\vct P^{\textrm{o}}_{l,n}}$, with polarization $P$~\cite{Tsang_Mie, trace}. Since both obey
$\mathbb{J}_z\ket{\vct P^{\alpha}_{l,n}}=\hbar n\ket{\vct P^{\alpha}_{l,n}}$ for $\alpha\in \{\textrm{r},\textrm{o}\}$, we have the matrix elements
\begin{align}
         {{\cal          T}_1}^{P,P'}_{l,l',n,n'}&=i\braket{\vct P^{\textrm{r}}_{ln}|\mathbb{T}_1|\vct P'^{\textrm{r}}_{l'n'}}=i\braket{\vct P^{\textrm{r}}_{ln}|\mathbb{T}_1|\vct P'^{\textrm{r}}_{l'n}}\delta_{n,n'}, \nonumber\\
         {{\cal          T}_2}^{P,P'}_{l,l',n,n'}&=i\braket{\vct P^{\textrm{o}}_{ln}|\mathbb{T}_2|\vct P'^{\textrm{o}}_{l'n'}}=i\braket{\vct P^{\textrm{o}}_{ln}|\mathbb{T}_2|\vct P'^{\textrm{o}}_{l'n}}\delta_{n,n'}.
    \end{align}
Using Lorentz reciprocity, i.e., $\mathbb{T}_{ij}(\mathbf{r},\mathbf{r}')=\mathbb{T}_{ji}(\mathbf{r}',\mathbf{r})$, we have, for either object,~\cite{Tsang_Mie, trace}
    \begin{align}
        {\cal T}(n) \equiv{{\cal T}}^{P,P'}_{l,l',n}&={{\cal T}}^{P',P}_{l',l,-n}={\cal T}^T(-n) .
    \end{align}
which yields ${\cal W}_{12}(-n)={\cal W}_{21}^T(n)$ and ${\cal R}_i(-n)={{\cal R}_i^*}^T(n)$. Furthermore, $\mathbb{A}_1=\mathbb{R}_1^*$ for reciprocal media~\cite{trace}. We thus have, in this case, 
\begin{align}
h_{-n}&=\mbox{Tr}\left[{\cal R}_1^T(n){\cal W}_{12}^T(n){{\cal R}_2^*}^T(n){{\cal W}_{21}^*}^T(n)\right] \nonumber \\ &=\mbox{Tr}\left[{\cal R}_2^*(n){\cal W}_{12}(n){{\cal R}_1}(n){{\cal W}_{21}^*(n)}\right] \nonumber \\&=h_n^*=h_n,
\end{align}
where in the last step we used that $h_n$ is real.
We therefore find that
\begin{align}
j=\frac{1}{\hbar}\mbox{Tr}[\mathbb{J}_z\mathbb{H}]=\sum_n nh_n=0,
\end{align}
which shows that for any two rotationally symmetric reciprocal objects, the torque vanishes. This statement is nontrivial, since rotationally symmetric objects include cases that break the chiral symmetry between clockwise and counterclockwise rotation, as depicted in Fig.~\ref{fig:Sketch2}.

\section{Explicit results for two cylinders}\label{sec:cyl}

From here on, we consider two concentric cylinders with homogeneous and isotropic dielectric responses, as depicted in Fig.~\ref{fig:setup}.

\subsection{Setup: Two concentric cylinders}
\begin{figure}[!htbp]
    \centering
    \includegraphics[width=\linewidth]{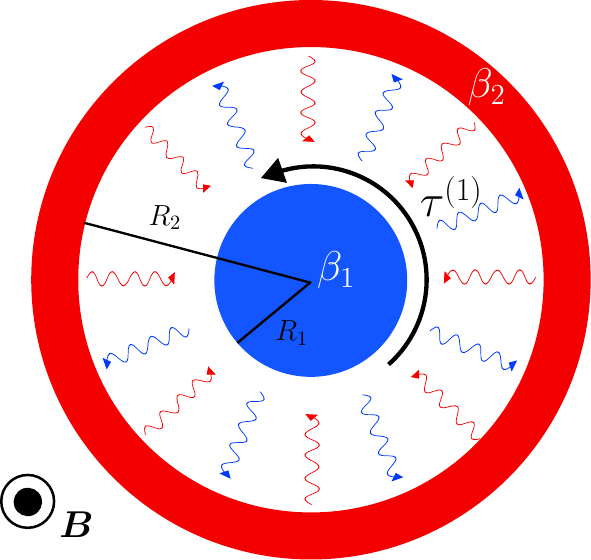}
    \caption{Cylindrical heat engine: Two concentric cylinders made of nonreciprocal material, each at different temperature. A magnetic field parallel to the cylinder axis breaks the symmetry of the cylinder's permittivity matrix, producing torques.
       }
    \label{fig:setup}
\end{figure}

We consider concentric cylinders separated by vacuum between radii $R_1$ and $R_2$, both with length $L \gg R_2$ and aligned with the $\hat z$-axis, as shown in Fig.~\ref{fig:setup}. The cylinders are kept at temperatures $T_1$ and $T_2$, respectively. The outer cylinder is considered much thicker than the skin depth of the material, so that the scenario introduced above of an enclosing geometry applies. The cylinders are made of non-magnetic materials with  
local dielectric tensors for cylinder $i \in \{1,2\}$ having the form
\begin{align}\label{eq:dielectric tensor}
    \eps_{i}(\w)
    = \eps_{d,i} \m I
    + (\eps_{p,i}-\eps_{d,i})\hat z \otimes \hat z
    + i\eps_{f,i} \hat z\times \ldots \, ,
\end{align}
where $\eps_{p,i}$, $\eps_{d,i}$, and $\eps_{f,i}$ are complex functions of
$\w$. The response is nonreciprocal if and only if $\eps_{f,i}$ is nonzero. nonreciprocity is achieved through an external magnetic field $B \hat z$~\cite{caloz_electromagnetic_2018,asadchy_tutorial_2020}, which is chosen to point along the cylinder axis, so that the setup remains rotationally symmetric around the $z$-axis and translationally invariant in the $\hat z$-direction.  As a result, $n$, the eigenvalue of $\mathbb{J}_z/\hbar $, and $k_z$, the wavenumber in the $z$ direction, are good quantum numbers.  While $\eps_{f,i}$ is odd in the magnetic field strength $B$, $\eps_{d,i}$ and $\eps_{p,i}$ are even functions of $B$. 

\subsection{Expansion in the cylindrical geometry}\label{sec: cyl wave basis}
We adopt cylindrical coordinates with the position vector $\vct{r}=(\rho, \phi,z)$ and use symmetries to express functions in cylindrical vector wavefunction basis~\cite{wengchochew, Tsang_Mie}.  The eigenfunctions of the Laplacian in cylindrical co-ordinates are given by $\psi_{n,k_z}^{\textrm{r}}(\vct r) \equiv J_n(k_\rho\rho)e^{in\phi+ik_zz}$, where $k_\rho = \sqrt{k^2-k_z^2}$, $J_n$ is the Bessel function of order $n$, and $(\n^2+k^2)\psi_{n,k_z}^{\textrm{r}} = 0$. The superscript $r$ indicates that the function is regular at the origin. Eigenfunctions $\psi_{n,k_z}^{\textrm{o}}(\vct r)$ obeying outgoing boundary conditions are defined by replacing the Bessel functions with $H_n^{(1)}(k_\rho \rho)$, the Hankel functions of the first kind. The three vector wavefunctions $L,M,N$ can be expressed as~\cite{wengchochew, Tsang_Mie}
\begin{align}\label{LMN functions}
    \vct L^\al_{n,k_z}(\vct r) &= \n \psi_{n,k_z}^\al(\vct r), \nonumber \\
    \vct M^\al_{n,k_z}(\vct r) &= \n \times \left(\hat z \psi_{n,k_z}^\al(\vct r) \right), \nonumber \\
    \vct N^\al_{n,k_z}(\vct r) &= \frac{1}{k} \n \times \n \times \left(\hat z \psi_{n,k_z}^\al(\vct r) \right) \,,
\end{align}
where $\al \in\{r,o\}$ denotes regular or outgoing, respectively. 
This basis diagonalizes the operator $\n\times \n\times(\ldots)$, and thus forms a complete set of vacuum solutions for the electric field, where only the transverse $M$ and $N$ polarized waves represent propagating modes. As mentioned before, the symmetries of the dielectric result in block diagonal scattering operators for the two cylinders, defined as
\begin{align}\label{eq: T operator to T matrix}
    \m T_{1}^{PP'}(n,k_z) \delta(k_z-k_z')\delta_{nn'}&=i\bra{\vct P_{n,k_z}^{\textrm{r}}}  \mb T_1   \ket{\vct P'^{\textrm{r}}_{n',k_z'}}  \, , \nonumber \\
    \m T_{2}^{PP'}(n,k_z) \delta(k_z-k_z')\delta_{nn'}&=i\bra{\vct P_{n,k_z}^{\textrm{o}}} \mb T_2   \ket{\vct P'^{\textrm{o}}_{n',k_z'}}  \, .
\end{align}
The scattering operators occur in most of our trace expressions in the combination $\mb G_0\mb T_i$, so we use the cylindrical wave expansion of the free Green function $\mb G_0$~\cite{wengchochew, rahi, trace} to obtain the equivalent relations,
\begin{align}\label{eq: G0T operator to T matrix}
    \m T_{1}^{PP'}(n,k_z) \delta(k_z-k_z')\delta_{nn'}&=\bra{\vct P_{n,k_z}^{\textrm{o}}}  \mb G_0\mb T_1   \ket{\vct P'^{\textrm{r}}_{n',k_z'}}  \, , \nonumber \\
    \m T_{2}^{PP'}(n,k_z) \delta(k_z-k_z')\delta_{nn'}&=\bra{\vct P_{n,k_z}^{\textrm{r}}} \mb G_0\mb T_2   \ket{\vct P'^{\textrm{o}}_{n',k_z'}}  \, .
\end{align}
As is implicit in the Dirac notation, we have $\bra{\vct P_{n,k_z}^\al}\vct r\rr\equiv \bra{\vct r}\vct P_{n,k_z}^\al\rr^* \equiv \vct P_{n,k_z}^{\al*}(\vct r)$. Note that this choice of basis and inner product is identical to that in Ref.~\cite{objectinsideobject}, but differs from Ref.~\cite{trace}. Ref.~\cite{trace} avoids the usage of complex conjugates $\vct P_{n,k_z}^*$, and replaces them with $\vct P_{-n,-k_z}$, which maintains the manifest analyticity of the Green functions $\mb G(\w)$ in the upper half complex plane $\text{Im} \,\w>0$. 
The matrices $\m T_1$ and $\m T_2$ can be computed by considering the cylindrical wave solution of the Maxwell equations inside and outside the cylinders and matching boundary conditions at the interface. Explicit expressions are included in Appendix \ref{sec reflection matrices}. For a reciprocal cylinder these are given, e.g., in Ref.~\cite{cylradiation}.

\subsection{Heat and angular momentum flux from outer to inner cylinder} 
As in Section \ref{sec:General}, we denote the heat transfer rate from object $i$ to $j$ as $H_i^{(j)}$, and the corresponding $z$-component of torque as $\tau_{i}^{(j)}$. Expanding Eqs.~\eqref{eq:heat trace} and \eqref{eq:torque general trace} in the cylindrical basis yields
\begin{align}\label{eq: cylinder hf torque}
    H_{2}^{(1)} &=
     4 \int_0^\infty  \frac{d\w}{2\pi}\nu_{\bt_2}(\w)  \sum_n \hbar \w\Phi_{2,n}^{(1)}(\w) \, \nonumber \\
    \tau_{2}^{(1)} &=  4 \int_0^\infty \frac{d\w}{2\pi}\nu_{\bt_2}(\w)  \sum_{n} \hbar n\Phi_{2,n}^{(1)}(\w) \, .
\end{align}
The heat flux density per mode $\Phi_{2,n}^{(1)}(\w)$ can be written explicitly in terms of a trace in the space of $2\times 2$ reflection matrices $\m T_1$ and $\m T_2$, with all matrices evaluated at $k_z$ and $n$, 
\begin{align}
    &\Phi_{2,n}^{(1)}(\w) = L\int\frac{dk_z}{2\pi } \text{Tr} \Biggl[ \left(\frac{\m T_2^\dagger+\m T_2}{2} + \Theta^\text{pr} \right) \frac{1}{1-\m T_1 \m T_2}\notag\\
    &\left( \frac{\m T_1 + \m T_1^\dagger}{2} + \Theta^\text{pr}\m T_1 \m T_1^\dagger\right) \frac{1}{1-\m T_2^\dagger \m T_1^\dagger}\Biggr] \, ,\label{hfd 12 in T matrix 2}
\end{align}
where $\Theta^\text{pr}\equiv \Theta(\w/c-|k_z|)$ projects onto propagating modes.

The expressions in Eq.~\eqref{eq: cylinder hf torque} show that the energy and angular momentum flux are determined by the corresponding photon energies $\hbar\omega$ and angular momenta $\hbar n$ weighted by $ \Phi_{2,n}^{(1)}(\w) $, which can be thought of as the probability density for a photon with a given energy and momentum to be emitted by cylinder 2 and absorbed by cylinder~1.

The above  formulas for heat flux and torque can be compared to the corresponding results for heat flux and tangential force reported in Ref.~\cite{nearfieldSI} for parallel plates oriented along the $yz$-plane. For parallel plates,  nonreciprocity was implemented by an external field along the $z$-axis, leading to a tangential force $F_y$ in the $\hat y$ direction. The expressions for heat flux and tangential force and torque in the two geometries are shown in Table \ref{hf table 2 geometries}. We immediately observe a correspondence between the azimuthal eigenvalues $n$ for the cylinder case and the translational eigenvalues $k_y$ in the parallel-plate case. Furthermore, the cylindrical heat flux density $\Phi_{2,n}^{(1)}$ in Eq.(\ref{hfd 12 in T matrix 2}) is analogous to $S_2^{(1)}(k_y)$, the heat flux density per wavevector for the translationally symmetric parallel plates~\cite{nearfieldSI}. Like $\Phi_{i,n}^{(j)}$, $S_i^{(j)}(k_y)$ can be explicitly written in terms of $2\times 2$ parallel-plate reflection matrices in polarization space~\cite{fanfan, nearfieldSI}. 

One can also derive the expressions in Eqs. (\ref{eq: cylinder hf torque}) and (\ref{hfd 12 in T matrix 2})  by explicitly integrating the Poynting vector and the electromagnetic stress tensor over a surface enclosing the inner cylinder. A detailed derivation is given in Appendix \ref{sec: hf poynting}. 

\begin{table*}[t]
    \caption{Heat flux and force expressions in the parallel-plate~\cite{nr} and cylinder geometry. We have defined $\nu_{\bt}'(\w)\equiv \p_\w \nu_{\bt}(\w)$. } 
    \begin{ruledtabular}
        \label{hf table 2 geometries}
        \begin{tabular}{ccc}
             & $\text{Parallel plates}$ &  $\text{Concentric cylinders}$\\
             \hline
             Heat flux  & $H_2^{(1)}=4\int_0^{\infty }\frac{d\w}{2\pi} \nu_{\bt_2}(\w) \int \frac{dk_y}{2\pi}\hbar\w S_{2}^{(1)}(k_y; \w)$ & $H_2^{(1)}=4\int_{0}^\infty\frac{d\w}{2\pi} \nu_{\bt_2}(\w) \sum_n\hbar\w \Phi_{2,n}^{(1)}(\w)$ \\
              Propulsive force/torque& $F_{y,2}^{(1)}=4\int_0^\infty \frac{d\w}{2\pi} \nu_{\bt_2}(\w)  \int \frac{dk_y}{2\pi} \hbar k_yS_{2}^{(1)}(k_y;\w)$ & 
              $\tau_2^{(1)}=4\int_{0}^\infty \frac{d\w}{2\pi}\nu_{\bt_2}(\w) \sum_n \hbar n\Phi_{2,n}^{(1)}(\w)$ \\
              Equilibrium friction & $\g_2^{(1)}=-4 \int_0^\infty \frac{d\w}{2\pi } \nu_{\bt_2}'(\w) \int \frac{dk_y}{2\pi }\hbar k_y^2 S_2^{(1)}(k_y;\w) $ & 
              $\g_2^{(1)}=-4\int_0^\infty \frac{d\w}{2\pi } \nu_{\bt_2}'(\w) \sum_n \hbar n^2 \Phi_{2,n}^{(1)}(\w) $
        \end{tabular}
    \end{ruledtabular}
\end{table*}


\subsection{Symmetries in pairwise heat flux densities}\label{hfd symmetries}
The net heat fluxes and forces are dependent on the four heat flux densities $\Phi_{i,n}^{(j)}$. As in the parallel-plate case~\cite{nearfieldSI}, there are additional relations between these quantities.
First, we can use direct trace manipulations in Eqs. (\ref{hfd 12 in T matrix}) and (\ref{hfd 22 in T matrix}) to give
\begin{align}\label{symrel1}
    \Phi_{1,n}^{(j)} = -\Phi_{2,n}^{(j)}\,, \, j \in \{ 1,2\},
\end{align}
which ensures that, in thermal equilibrium,  the inner cylinder is neither heated nor cooled. Another symmetry relation exists in our geometry with one object enclosed in another. Due to the contour independence of the integral of the Poynting vector, the heat flux due to fluctuations from object $i$ through object 1 is the negative of that through object 2, with the minus sign arising because the normal vector $\hat n$ flips,
\begin{align}\label{symrel2}
    \Phi_{i,n}^{(1)} = - \Phi_{i,n}^{(2)}\,, \, i \in \{ 1,2\}.
\end{align}
We can also verify this equality explicitly by using Eqs.~(\ref{hfd 12 in T matrix}) and (\ref{hfd 22 in T matrix}). Note that this proof uses only the cyclic properties of the trace, which means that it does not require an explicit expansion of the heat flux density in the cylindrical basis. The trace manipulations can be carried out directly at the level of operators. 

Using Eqs.~\eqref{symrel1} and  \eqref{symrel2}, we get the following expressions for the net heat flux $H^{(j)}\equiv H_1^{(j)}+H_2^{(j)}$, and net torque experienced by the two cylinders $\tau^{(j)}\equiv \tau_1^{(j)}+\tau_2^{(j)}$,
\begin{align}
    H^{(1)}&=-H^{(2)}
     =4 \int_0^\infty  \frac{d\w}{2\pi}\left[\nu_{\bt_2} - \nu_{\bt_1} \right] \sum_n \hbar \w\Phi_{2,n}^{(1)}(\w) \, \nonumber \\
    \tau^{(1)} &= -\tau^{(2)} = 4 \int_0^\infty \frac{d\w}{2\pi}\left[\nu_{\bt_2} - \nu_{\bt_1} \right] \sum_{n} \hbar n\Phi_{2,n}^{(1)}(\w) \, .\label{eq:handt}
\end{align}
We see that in thermal equilibrium, $\bt_1 = \bt_2$, there is no net heat flux or propulsive torque, and thus no work can be extracted from the engine, analogously to the same statement in the Feynman ratchet problem~\cite{Feynman1963}. Furthermore, the net heat emitted by the outer cylinder is exactly that absorbed by the inner one, since there are no other sources or sinks, and the forces obey Newton's Third Law even though the materials are nonreciprocal. Both findings are consequences of the absence of the environment as a ``third object,'' which, if present, may take up energy or angular momentum.

\subsection{Explicit torque cancellation in the reciprocal limit}
\label{sec: small eta torque cyl}
We have shown in Sec.~\ref{sec:zero} above that in a system with two rotationally symmetric reciprocal objects, the propulsive torque is zero. Taking the special case of two cylinders and our spatially local dielectric, we can check this result explicitly. On taking the reciprocal limit $\eps_f \rightarrow0$ in the dielectric tensor in Eq.~\eqref{eq:dielectric tensor}, we find that the reflection matrices for the cylinders, computed in Appendix \ref{sec reflection matrices},  obey the relations, for $i \in \{ 1,2\}$,
\begin{align}\label{mode cancellation}
    \m T_i(n,k_z) = \m T^T_i(-n,-k_z)= \m T_i(-n,-k_z)= \m T^T_i(n,k_z),
\end{align}
which yields $\Phi_{i,n}^{(j)} = \Phi_{i,-n}^{(j)}$, thus implying that all torques $ \tau_i^{(j)} $ are zero, as is the net torque in Eq.~\eqref{eq:handt}. Physically, this result implies a cancellation between transverse photons carrying angular momenta $+n$ and $-n$ between the cylinders. In the weak nonreciprocity limit $\eps_f \ll \eps_d$, one can show that the symmetry-breaking term in Eq.~\eqref{mode cancellation} is linear in $\eps_f$, giving $\tau^{(i)} \propto \eps_f$ for small $\eps_f$. 

The first equality in Eq.~\eqref{mode cancellation} is the fundamental symmetry for reciprocal media \cite{Tsang_Mie,trace}, while second and third equalities are a consequence of the symmetry of the permittivity in Eq.~\eqref{eq:dielectric tensor} under reflection with respect to the cylinder axis.
\subsection{A thin nonreciprocal cylinder exchanging heat and angular momentum with environment}

A special case of the setup above is a single nonreciprocal cylinder at temperature $\bt_1$ placed in an  environment of temperature $\bt_{\text{e}} \neq \bt_1$. In this case, 
heat transfer and torque are obtained by letting $\eps_2\to1$, i.e., by letting the reflection matrix $\m T_2\to0$ in Eq.\ (\ref{hfd 12 in T matrix 2}) with $\beta_2\to\beta_{\text{e}}$, which yields
\begin{align}\label{eq: 1 cyl torque heat}
    H_{(\text{e})}^{(1)} &=4 \int_0^\infty \frac{d\w}{2\pi} \left[\nu_{\bt_{\text{e}}}-\nu_{\bt_1} \right] \sum_n \hbar \w \, \Phi_{\text{e},n}^{(1)} \nonumber \\
    \tau_{(\text{e})}^{(1)} &=4 \int_0^\infty \frac{d\w}{2\pi} \left[\nu_{\bt_{\text{e}}}-\nu_{\bt_1} \right]\sum_n \hbar n \, \Phi_{\text{e},n}^{(1)} \,.
\end{align}
The heat flux density $\Phi_{\text{e},n}^{(1)}$ between the environment and the cylinder then becomes 
\begin{align}\label{eq: 1 cyl hfd go}
    \Phi_{\text{e},n}^{(1)} &\equiv L\int_{-\w/c}^{\w/c} \frac{dk_z}{2\pi } \sum_{P} \left[  \text{Re} \left( \m T_{n,k_z}^{PP} \right)+ \sum_Q|\m T_{n,k_z}^{PQ}|^2 \right]\,,
\end{align}
where, in this subsection, we  omit the object index on $\m T$ and $\eps$. 

The heat radiation from a reciprocal cylinder was studied in Refs.~\cite{shorttrace,cylradiation}. 
To consider the nonreciprocal case, we expand the exact result for $\m T$ given in Eq.~\eqref{eq:T1} to lowest order in nonreciprocity,  by keeping only the lowest order in $\eps_f$ and setting $\eps_p = \eps_d$, since $\eps_p-\eps_d \simeq \m O(B^2)$. 

As shown in  Appendices \ref{sec reflection matrices} and \ref{app: thin cyl reften}, the reflection coefficients $\m T_{n,k_z}^{PP'}$, and the transmission coefficients $A_{n,k_z}^{ \pm P'}$ are given by the eight linear equations
\begin{align}\label{eq: 8 linear eqs} 
    \begin{pmatrix}
        \m T^{MM}\\ \m T^{NM}\\ A^{+M}\\ A^{-M} 
    \end{pmatrix} 
    &= \m K_n^{-1}(\eps_f) \begin{pmatrix}
        -k_\rho J_n'(k_\rho R) \\
        0 \\
        \frac{nk_zc}{\w R}J_n(k_\rho R) \\
        \frac{ck_\rho^2}{\w} J_n(k_\rho R)
    \end{pmatrix} \, , \nonumber \\
    \begin{pmatrix}
        \m T^{MN}\\ \m T^{NN}\\ A^{+N}\\ A^{-N} 
    \end{pmatrix} 
    &= \m K_n^{-1}(\eps_f) \begin{pmatrix}
        \frac{nk_zc}{\w R}J_n(k_\rho R) \\
        \frac{ck_\rho^2}{\w} J_n(k_\rho R) \\
        -k_\rho J_n'(k_\rho R) \\
        0 
    \end{pmatrix} \,,
\end{align}
where we have defined $k_\rho \equiv \sqrt{\w^2/c^2-k_z^2}$. The matrix $\m K_n(\eps_f)$ can be expanded as $\m K_n(\eps_f)=\sum_{r=0}^\infty\eps_f^r \m K_{n,(r)}$, so that $\m K_n^{-1} = \m K_{n,(0)}^{-1}- \eps_f \m K_{n,(0)}^{-1} \m K_{n,(1)} \m K_{n,(0)}^{-1} + \m O(\eps_f^2) $.

We can use the Bessel function identity $J_{-n}(z)=(-1)^n J_n(z)$ to obtain relations between the matrices $\m K_{n,(r)}$ and $\m K_{-n,(r)}$. In particular,
\begin{align}\label{eq: n=0,1 K symmetries}
    \m K_{-n,(0)} &= \m L\m K_{n,(0)}\m R
    \quad \hbox{and} \quad
    \m K_{-n,(1)} = -\m L \m K_{n,(1)} \m R\,,
\end{align}
with
\begin{align}
    \m L &\equiv \text{diag (-1,1,1,-1)}\,,\, \m R\equiv \begin{pmatrix}
        1&0&0&0 \\
        0&-1&0&0\\
        0&0&0&1\\
        0&0&1&0
    \end{pmatrix}\,,
\end{align}
which guarantees that in the reciprocal case, where $\m K = \m K_{(0)}$, we have $\m T_n^{PP'} = (2\delta^{PP'}-1)\, \m T_{-n}^{PP'}$, and $\Phi_{n}=\Phi_{-n}$, such that the net torque vanishes. The opposite sign for $\m K_{n,(1)}$ in Eq.~\eqref{eq: n=0,1 K symmetries} implies that the symmetry-breaking term for the torque is of order $\eps_f$. Explicit expressions for $\m K_{n,(0)}$ and $\m K_{n,(1)}$ are given in Eq.~\eqref{K0 K1 expressions}.

It is instructive to take the thin cylinder limit, i.e., $R \equiv R_1 \ll  \delta$ and $\lambda_T$, where $\delta$ is the skin depth and $\lambda_T$ is the relevant thermal wavelength for the dielectric response at temperatures $\bt_1$ and $\bt_{\text{e}}$~\cite{cylradiation}. In this case, modes with $|n| \geq 2$ are negligible, as shown in Appendix \ref{app: thin cyl reften}, and  only  reflection matrix elements for $n = 0$ and $n=\pm 1$ are needed to lowest orders in $R\w/c$ and $\eps_f$. For $n=0$, the only component contributing at order $R^2$ is
\begin{align}\label{eq: n=0 oef2}
    &\m T_{0,k_z}^{NN} = \frac{i\pi k_\rho^2R^2}{4}(\eps_d-1) + \m O(\eps_f^2). 
\end{align}
For $n = \pm 1$, we get
\begin{align}\label{eq: n = 1 oef2}
    \m T_{\pm 1, k_z}^{MM} &= \frac{i\pi R^2\w^2}{4c^2} \left[ \frac{\eps_d-1}{\eps_d+1} \mp \frac{\eps_f}{(\eps_d+1)^2}  \right] \nonumber \\
    \m T_{\pm 1, k_z}^{NN} &= \frac{i\pi R^2k_z^2}{4} \left[ \frac{\eps_d-1}{\eps_d+1} \mp \frac{\eps_f}{(\eps_d+1)^2}  \right] \nonumber \\
    \m T_{\pm 1, k_z}^{MN} &= \pm\frac{i\pi R^2\w k_z}{4c} \left[  \frac{\eps_d-1}{\eps_d+1} \mp \frac{\eps_f}{(\eps_d+1)^2}  \right] \nonumber \\
    \m T_{\pm 1, k_z}^{NM} &= \pm\frac{i\pi R^2\w k_z}{4c} \left[  \frac{\eps_d-1}{\eps_d+1} \mp \frac{\eps_f}{(\eps_d+1)^2}  \right] .
\end{align}
Details of these calculations are given in Appendix \ref{app: thin cyl reften}. Setting $\eps_f = 0$ in Eqs.~\eqref{eq: n=0 oef2} and \eqref{eq: n = 1 oef2}, we recover the reciprocal reflection matrix, as derived in Ref.~\cite{cylradiation}.

Finally, using the $\m T$-matrix elements in Eqs.~\eqref{eq: 1 cyl torque heat} and \eqref{eq: 1 cyl hfd go}, we get
\begin{align}
    \frac{H_{(\text{e})}^{(1)}}{L}&= \frac{2R^2\hbar}{3c^3}\int_0^\infty \frac{d\w}{2\pi } \w^4 \,\Delta\nu(\w) \, \text{Im} \left[ \frac{(\eps_d-1)(\eps_d+5)}{\eps_d+1} \right] \nonumber, \\
    \frac{\tau_{(\text{e})}^{(1)}}{L} &= -\frac{8R^2 \hbar }{3c^3} \int_0^\infty \frac{d\w}{2\pi} \w^3 \Delta \nu (\w)\, \text{Im} \left[\frac{\eps_f}{(\eps_d+1)^2} \right]\,,\label{eq:heat_torque}
\end{align}
where we have defined $\Delta \nu(\w) \equiv \nu_{\bt_{\text{e}}}(\w)-\nu_{\bt_1}(\w)$. We explicitly see that the torque is proportional to the nonreciprocity coefficient $\eps_f$, while the heat flux is modified only at $\m O(\eps_f^2)$. 

Equation~\eqref{eq:heat_torque} may give the impression that the torque can be arbitrarily large compared to the heat transfer for $\eps_d\to1$ with $\eps_f$ fixed, which would violate the bound in Eq.~\eqref{eq:bound}. However, $\eps_d\to1$ with $\eps_f$ fixed violates $(\mathbb{V}-\mathbb{V}^\dagger)/2i\geq 0$, which is a condition for a passive medium, as is required for the bound of Eq.~\eqref{eq:bound} to hold.

\section{Stability of the inner cylinder}\label{sec:stab}
We next study whether it is possible to stably levitate a (possibly rotating) inner cylinder without mechanical connections.
To determine the stability of the setup in Fig.~\ref{fig:setup}, we consider the force on the inner cylinder under small displacements from the $z$-axis. We address this question within the proximity force approximation, first finding the appropriate stability constraints and then finding conditions that fulfill these constraints.

\subsection{Proximity approximation}\label{subsec:Proximity}
Let $\rho_{\textrm{c}}$ indicate the deviation of the inner cylinder from the central axis. We define an energy $E(\rho_{\textrm{c}})=E_0+\int_0^{\rho_c}d\rho F_\rho$, with $F_\rho$ the radial force acting on the inner cylinder. By symmetry,  $F_\rho(\rho_c=0)=0$ and $E(\rho_c)=E_0+\kappa\rho_c^2+\dots$ for small $\rho_{\textrm{c}}$, where $\kappa>0$ corresponds to stable equilibrium. 

\begin{figure}[!htbp]
\begin{center}
\includegraphics[width=1.0\linewidth]{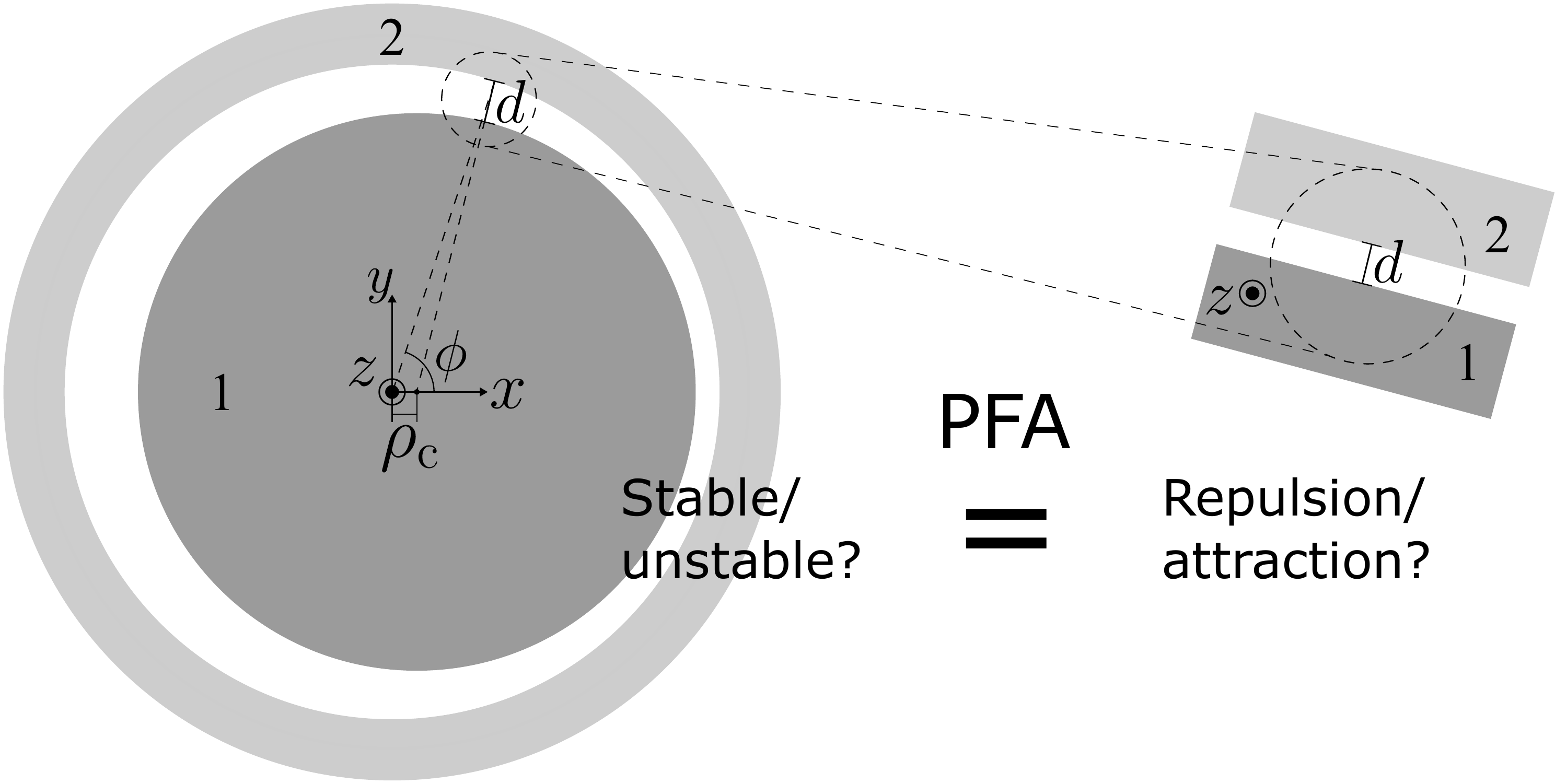}
\end{center}
\caption{\label{fig:PFA}The inner cylinder is slightly displaced by distance $\rho_{\textrm{c}}$ along the $x$ axis. The cylinder is stable (i.e., it returns to the origin) if the force between the cylinder surfaces at angle $\phi$ is repulsive and decays with the surface-to-surface distance $d$. In the proximity force approximation (PFA), this requirement is equivalent to having repulsion between parallel plates made of the same materials.}
\end{figure}

For parallel plates at separation $d$, the energy per unit area is denoted by ${\cal E}_{\textrm{pp}}$. Within PFA, the energy for arbitrary surfaces is obtained by considering small area elements $dA$ facing each other at separation $d$, with
\begin{align}
dE={\cal E}_{\textrm{pp}}(d)d{A}.
\end{align}
For two cylinders with $R_2-R_1\ll R_1$, with the inner cylinder  displaced from the $z$-axis along $x$ as shown in Fig.~\ref{fig:PFA},  the distance $d$ between the cylinder surfaces at an angle $\phi$ is 
\begin{align}
   \notag & d(\phi)=\sqrt{R_2^2+\rho_{\textrm{c}}^2-2R_2\rho_{\textrm{c}}\cos\phi}-R_1 = (R_2-R_1)\\
   & \times\left(1-r\cos(\phi)+\frac{r^2}{2}\frac{R_2-R_1}{R_2}\sin^2(\phi)+\dots\right),
\end{align}
with $r\equiv \rho_{\textrm{c}}/(R_2-R_1)$.
Integrating over angle gives
\begin{align}
\frac{E(\rho_{\textrm{c}})}{R_1L}=\int_0^{2\pi}d\phi \, {\cal E}_{\textrm{pp}}(d(\phi)).
\end{align}
We assume that ${\cal E}_{\textrm{pp}}$ scales as a power law in distance $d$ (as is the case for Casimir and van der Waals interactions), to obtain
\begin{align}
\notag {\cal E}_{\textrm{pp}}(d)&={\cal E}_0\frac{(R_2-R_1)^\alpha}{d^\alpha}\\ \notag &={\cal E}_0\biggl(1+\alpha r\cos\phi+\frac{r^2}{2}\alpha(\alpha+1)\cos^2\phi\notag\\& \ \ \ \ \ \ \ \ \ \ \,-\frac{\alpha r^2}{2}\frac{R_2-R_1}{R_2}\sin^2(\phi)+\dots\biggr)\label{eq:PFA}.
\end{align}
Neglecting the last term since $R_2-R_1\ll R_2$, integration over angle yields
\begin{align}
\frac{E}{2\pi R_1L}={\cal E}_0\left(1+\frac{1}{4} \alpha(\alpha+1)r^2 +\dots\right).
\end{align}
We thus find that the center position is stable against small perturbations if
\begin{align}
    {\cal E}_0\alpha(\alpha+1)>0.
\end{align}
From Eq.~\eqref{eq:PFA}, the force between parallel plates per unit area, i.e., the pressure $P$, is
\begin{align}
    P=-\frac{\partial}{\partial d}{\cal E}_{\textrm{pp}}(d)={\cal E}_0\alpha\frac{(R_2-R_1)^\alpha}{d^{\alpha-1}}.
\end{align}
Assuming $\alpha>0$, so that the energy decays in magnitude with increasing distance, we find that if $P<0$, corresponding to an attractive force, the configuration is unstable, while 
$P>0$, corresponding to a repulsive force, it is stable.

We have thus shown that within PFA, a repulsive force between parallel plates yields a stable configuration when the cylinder axes coincide. In the next subsection we analyze the possibility of repulsion for reciprocal and nonreciprocal parallel plates at different temperatures.

\subsection{Repulsion between closely spaced nonreciprocal parallel plates}
\label{subsec:PlateRepulsion}
Consider two semi-infinite nonreciprocal parallel plates, separated by a distance $d$ and held at inverse temperatures $\beta_1$ and $\beta_2$. Their permittivity tensors $\bs \eps_1$ and $\bs \eps_2$ are given by Eq.~\eqref{eq:dielectric tensor}, with the external magnetic field applied in a direction parallel to the surface (corresponding to the $z$-axis for cylinders). We are interested in the Casimir pressure $P^{(1)}$ acting on plate 1, in order to assess the possibility of repulsion in the parallel-plate geometry. For reciprocal nonmagnetic plates in vacuum, the force is generally attractive~\cite{PhysRevLett.97.160401,Bordag2009, Rahi2010}. Achieving repulsion therefore requires additional ingredients, such as immersion in a liquid~\cite{Bordag2009}, or  exotic material properties~\cite{Zhao2009, Grushin2011, Castillo-Lopez2022}. Nonequilibrium conditions offer another route: repulsive forces have been found for dilute plates at different temperatures with tuned resonances~\cite{Bimonte2011}. Here, we extend this scenario to nonreciprocal plates and formulate corresponding requirements.

The total pressure can be separated into three components~\cite{trace},
\begin{equation}
P^{(1)}(\beta_1,\beta_2)= P_0^{(1)} + P_1^{(1)}(\beta_1) + P_2^{(1)}(\beta_2),
\label{eq:P1separation}
\end{equation}
where $P_0^{(1)}$ is the equilibrium zero-point pressure, $P_1^{(1)}(\beta_1)$ is the thermal self-pressure due to sources in plate 1, and $ P_2^{(1)}(\beta_2) $ is the thermal interaction pressure due to sources in plate 2. Since we neglect any pressure acting on plate 1 from the outside environment, we naturally have $P_2^{(1)}(\beta_2)= -P_2^{(2)}(\beta_2)$, so that $P_2^{(1)}(\beta_2)$ may be found from $P_1^{(1)}(\beta_1)$ by exchanging all plate indices.   

General expressions for the three terms in Eq.~\eqref{eq:P1separation} are known (see, for example, Refs.~\cite{Zhao2009, Grushin2011} for $P_0^{(1)}$ and Refs.~\cite{nr,trace,Iizuka2023} for $P_i^{(1)}$). Here, we restrict to small distances, namely 
$ d \ll 2\pi c/\omega$ and
$ d \ll c/\Big(\omega\sqrt{|\epsilon_{\{p,d,f\},i}(\omega)|}\Big)$ in the range of relevant frequencies. In this case, contribution of propagating waves to the force can be neglected, and only the diagonal electric-electric element $ r_i^{NN} $ of the Fresnel matrix is relevant~\cite{nr}. Each component scales as $ d^{-3} $ and can be expressed via the trilogarithm $\textrm{Li}_3$,
\begin{equation}
P_0^{(1)} = -\frac{\hbar}{8\pi^3d^3}\int_0^{\infty}d\omega \, \frac{1}{2}\int_0^{2\pi}d\theta \, \textrm{Im}\left[\textrm{Li}_3\left(r_1^{NN}r_2^{NN}\right)\right],
\label{eq:P0}
\end{equation}
\begin{align}
\notag P_i^{(1)}(\beta_i) = & -\frac{\hbar}{8\pi^3 d^3}\int_0^{\infty} d\omega \,\nu_{\beta_i}(\omega)\int_0^{2\pi}d\theta\\
& \times \textrm{Im}\left[\textrm{Li}_3\left(r_1^{NN}r_2^{NN}\right)\right]\frac{\textrm{Im}\left[r_i^{NN}\right]\textrm{Re}\left[r_{\overline{i}}^{NN}\right]}{\textrm{Im}\left[r_1^{NN}r_2^{NN}\right]},
\label{eq:Pi1}
\end{align}
where $ \theta $ is the in-plane angle of the wave vector. A positive (negative) pressure corresponds to repulsion (attraction). For small $d$, the Fresnel coefficient can be approximated as~\cite{nr}
\begin{equation}
r_i^{NN}(\omega,\theta) = \frac{\epsilon_{d,i}(\omega)-1+\eps_{f,i}(\omega)\sin\theta}{\epsilon_{d,i}(\omega)+1+\eps_{f,i}(\omega)\sin\theta}.
\label{eq:rNN}
\end{equation}
For reciprocal plates, $ \eps_{f,i} = 0 $ and $ r_i^{NN} $ is independent of $ \theta $, so the angular integral in Eqs.~\eqref{eq:P0} and~\eqref{eq:Pi1} gives $2\pi$, recovering the expressions in Refs.~\cite{Antezza2008, Bordag2009, Bimonte2011}. In equilibrium ($\beta_1=\beta_2\equiv\beta$), adding the self and interaction pressures in Eq.~\eqref{eq:Pi1} leads to Eq.~\eqref{eq:P0} with $1/2$ replaced by $\nu_{\beta}$.

Equations~\eqref{eq:P0}--\eqref{eq:rNN} allow us to explore the possibility for repulsion analytically. To do so, we approximate $ \textrm{Li}_3\left(r_1^{NN}r_2^{NN}\right) \approx r_1^{NN}r_2^{NN} $, which is applicable in the case of dilute plates with small $ r_i^{NN} $. Considering the Drude-Lorentz model for the dielectric response and restricting to small external magnetic field, the imaginary and real parts of the Fresnel coefficient can be written as in Eqs.~\eqref{eq:rNN_DL_Im} and~\eqref{eq:rNN_DL_Re}, respectively. Using these approximations in Eqs.~\eqref{eq:P0} and~\eqref{eq:Pi1}, and assuming $|\omega_1-\omega_2| \gg \gamma $ and $ \hbar\omega_i\beta_i \ll 1$, where $\omega_i$ is the resonance frequency of the Fresnel coefficient for $B = 0$ and $ \gamma \ll \omega_i$ is the damping rate, in Appendix~\ref{app:DL} we obtain for the sum of all pressures
\begin{align}
\notag P^{(1)}(\beta_1,\beta_2) & \approx \frac{1}{32\pi d^3\beta_2}\left(\frac{\omega_{\textrm{p}}}{\omega_2}\right)^4\frac{1}{\widetilde{\omega}^2\left(\widetilde{\omega}+1\right)}\\
& \times \left[-\frac{\hbar\omega_2\beta_2}{2}\widetilde{\omega} + \frac{\widetilde{\beta}-\widetilde{\omega}^2}{\widetilde{\omega}-1}\right] + \mathcal{O}\left(B^2\right),
\label{eq:P1an}
\end{align}
where $ \widetilde{\beta} \equiv \beta_2/\beta_1 $, $ \widetilde{\omega} \equiv \omega_1/\omega_2 $, and $ \omega_{\textrm{p}} $ is the plasma frequency. Note that $ \left|\widetilde{\omega}-1\right| \gg \gamma/\omega_2 $, i.e., $ \widetilde{\omega} $ cannot be arbitrarily close to unity for Eq.~\eqref{eq:P1an} to hold. A comparison between Eq.~\eqref{eq:P1an} and exact numerical results is given in Fig.~\ref{fig:Pressure_plates}(b).

\begin{figure}[!htbp]
\begin{center}
\includegraphics[width=1.0\linewidth]{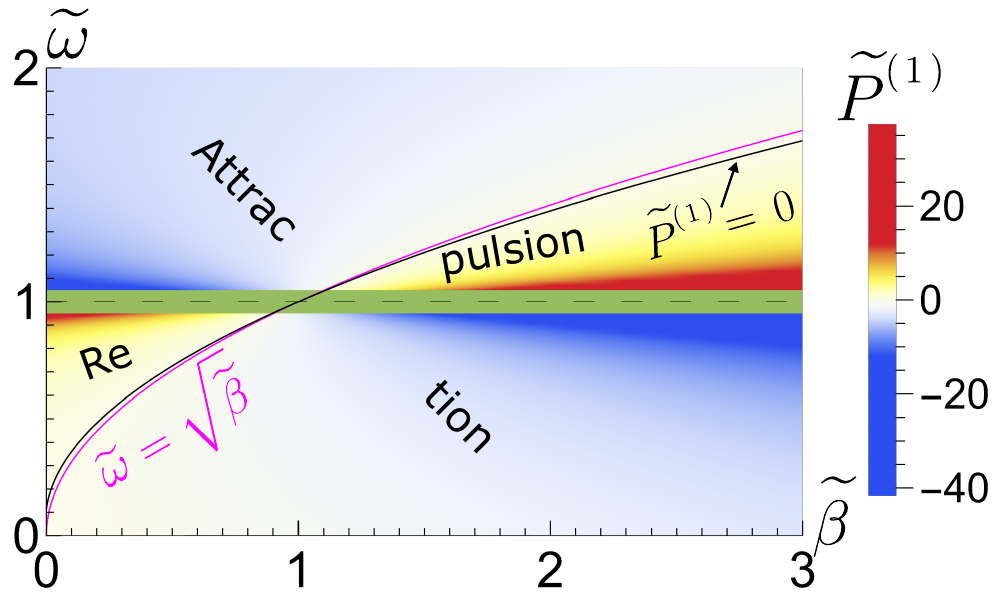}
\end{center}
\caption{\label{fig:RepulsionAttraction}Repulsive ($ \widetilde{P}^{(1)} > 0 $) and attractive ($ \widetilde{P}^{(1)} < 0 $) regimes for $ P^{(1)} $ according to Eq.~\eqref{eq:P1an}, with $ \hbar\omega_2\beta_2 = 0.26 $.  Here $ \widetilde{P}^{(1)} $ denotes the bracketed contribution in Eq.~\eqref{eq:P1an}. The green stripe indicates the range where Eq.~\eqref{eq:P1an} is not accurate, because $ \left|\widetilde{\omega}-1\right| \gg \gamma/\omega_2 $ is not satisfied;  we use $\gamma/\omega_2=0.01$. The black solid curve marks zero pressure, while the magenta solid curve shows $ \widetilde{\omega} = \sqrt{\widetilde{\beta}} $. A horizontal dashed line at  $ 1 $ is included to guide the eye.}
\end{figure}


The zero-point contribution to $ P^{(1)} $ in Eq.~\eqref{eq:P1an}, given by the first term in the square brackets, is always negative (i.e., attractive)\footnote{The same follows from Eq.~\eqref{eq:P1an} for the equilibrium pressure at finite temperature.}. The thermal part,  given by the second term, can have either sign. It is positive (i.e., repulsive) if $ 1 < \widetilde{\omega} < \sqrt{\widetilde{\beta}} $ or $ \sqrt{\widetilde{\beta}} < \widetilde{\omega} < 1 $. As depicted in Fig.~\ref{fig:RepulsionAttraction}, this condition accurately sets the repulsive regime for $ P^{(1)} $, because the thermal part in Eq.~\eqref{eq:P1an} dominates over the zero-point part unless $ \widetilde{\omega} \approx \sqrt{\widetilde{\beta}} $. In other words, $ P^{(1)} $ is repulsive if (i) the resonance frequency of the hotter plate is larger than that of the colder one, and (ii) their ratio is smaller than the square root of the ratio of the temperatures. Nonreciprocity enters the pressure to order $B^2$, so that a small value of $ B $ does not affect this discussion. We find numerically that this statement remains true even for larger values of $B$, as demonstrated in Fig.~\ref{fig:Pressure_plates}(a). As expected, since the magnetic field is parallel to the plates, the normal force is even in $B$ and the tangential force~\cite{nr} is linear in $B$ for small $B$.

\section{Slowly rotating inner cylinder, friction and engine efficiency}\label{sec:rot}
\begin{figure}[!htbp]
    \centering
    \includegraphics[width=0.8\linewidth]{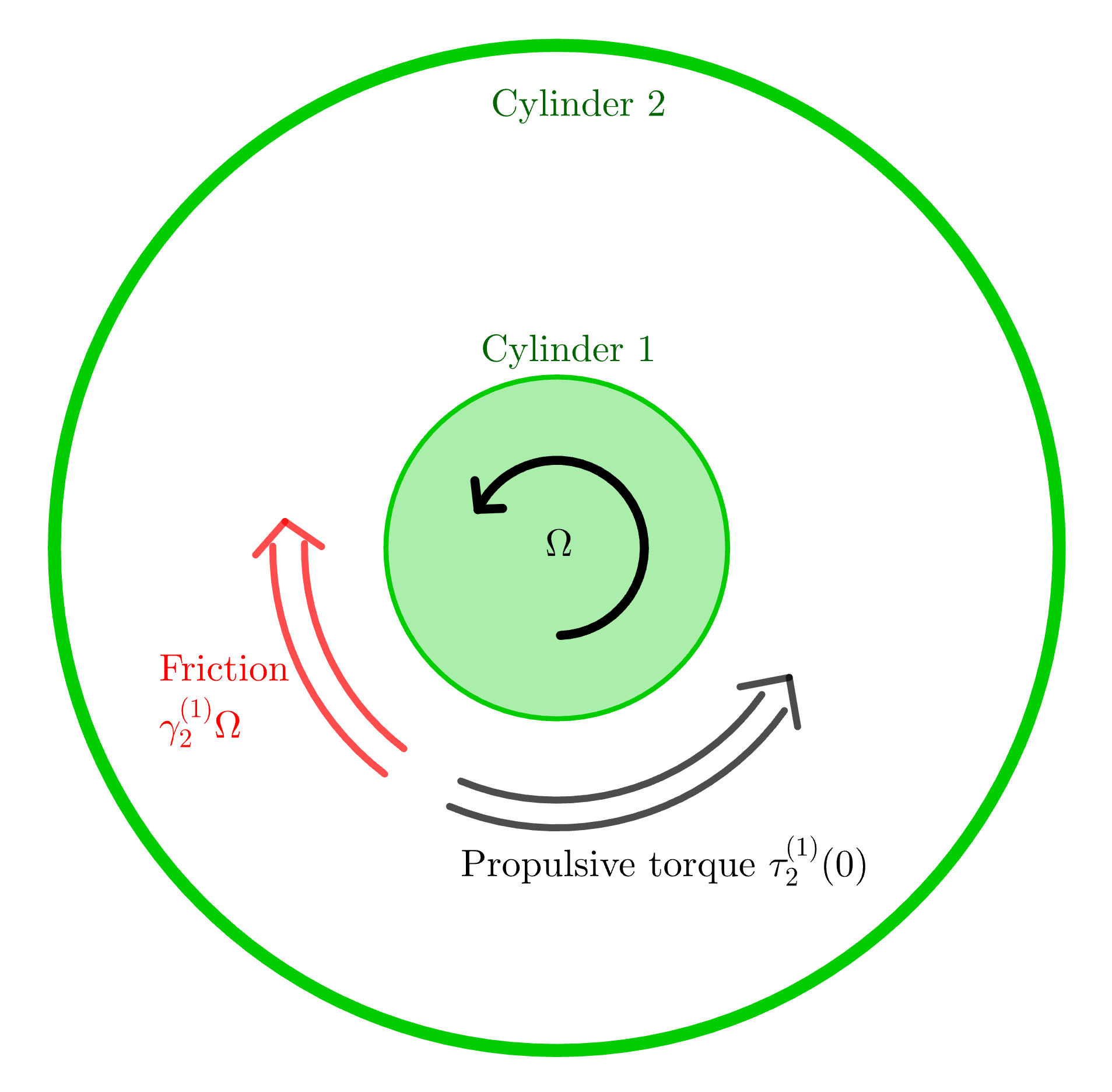}
    \caption{Propulsive torque and friction acting on the inner rotating cylinder.}
    \label{friction 2cyl}
\end{figure}

To extract work from the engine setup of Fig.~\ref{fig:setup}, it is now assumed that the inner cylinder rotates with angular frequency $\W$ around the $z$-axis, as shown in Fig.~\ref{friction 2cyl}.

\subsection{Heat flux and torque}
Both heat flux and torque are modified when $\W$ is nonzero.  Expanding to first order in $\W$, we have
\begin{align}
    H_i^{(j)}(\W) &= H_i^{(j)}(0)+\W h_i^{(j)}+\m O(\W^2),\nonumber \\ \tau_{i}^{(j)}(\W)&= \tau_{i}^{(j)}(0)-\g_i^{(j)}\W+\m O(\W^2).
\end{align}
To obtain the engine efficiency it is necessary to calculate the linear response functions $h_i^{(j)}$ and $\g_i^{(j)}$, where the latter represents the friction coefficient. To do so, we need the electric field correlator in the case of a rotating cylinder. We find that this correlator, like the heat transfer and torque, requires two ingredients. The first is the transformed potential  $\mb V_1^\W$ of the inner cylinder as seen from the rest frame, which can be found using the Lagrangian~\cite{movingcorr}. In general, it may not be stationary, i.e. diagonal in $\w$, but for the case of  azimuthal symmetry, $\mb V_1^\W$ is diagonal in both $n$ and $\w$. Indeed, $\bra{\vct r}\mb V_1^{\W} \ket{\vct r'}\equiv \mb V_1^\W(\rho,\rho';z,z';\phi-\phi')$, so we may define the angular-momentum-resolved potential through  $\delta_{nn'}\m V_{1,n}^{\W}(\rho,\rho';z,z')=\int_0^{2\pi}\int_0^{2\pi}\frac{d\phi \, d\phi'}{(2\pi)^2}e^{in\phi-in'\phi'}\bra{\vct r}\mb V_1^\W\ket{\vct r'} $. We can then show that
\begin{align}\label{eq: V rot}
    \m V_{1,n}^\W(\w) &= \frac{\w^2}{c^2} \m D_{n}^\dagger \left[ \bs \eps_1(\w-n\W)-1 \right] \m D_{n} \nonumber \, ,\\
    \mb D &\equiv \mb I + \frac{\rho\W}{i\w} \hat \phi \times \n \times (\ldots) + \m O(\W^2)\,,
\end{align}
where the Lorentz transformation operator $\mb D$ is also diagonal in $n$, and $\m D_{n}(\rho,\rho';z,z') \delta_{nn'}\equiv \int_0^{2\pi}\int_0^{2\pi}\frac{d\phi d\phi'}{(2\pi)^2}e^{in\phi-in'\phi'}\bra{\vct r}\mb D \ket{\vct r'}$. 

The transformed potential gives rise to the transformed scattering operator $\mb T_{1}^\W \equiv \mb V_1^\W \left[1-\mb G_0 \mb V_1^\W \right]^{-1}$, and all previously established trace formulae~\cite{trace} hold with the replacement of the static scattering operator $\mb T_1$ by the transformed one, $\mb T_1^\W$. In particular, the pairwise heat flux symmetries from Sec.~\ref{hfd symmetries} still hold. 

The second consequence of rotation is that the Bose factor for sources in the inner cylinder is evaluated at the shifted frequency $\w-n\W$~\cite{movingcorr}. The total heat flux to and the torque on the inner cylinder are thus
\begin{align}\label{total rotating torque}
    H^{(1)} &= 4\int_0^\infty \frac{d\w}{2\pi} \sum_n\left[\nu_{\bt_2}(\w) - \nu_{\bt_1}(\w-n\W) \right] \, \hbar \w\Phi_{2,n}^{(1), \W} \, , \nonumber \\
    \tau^{(1)} &= 4\int_0^\infty \frac{d\w}{2\pi} \sum_n\left[\nu_{\bt_2}(\w) - \nu_{\bt_1}(\w-n\W) \right] \, \hbar n\Phi_{2,n}^{(1), \W} \, .
\end{align}
Here $\Phi_{2,n}^{(1),\W}(\w)$ is the heat flux density at finite $\Omega$, with the same trace formula as the static case up to the replacement $\mb T_1 \rightarrow \mb T_1^\W$. 

We can consider the simpler limit of a small temperature difference $0\leq \Delta\bt \equiv \bt_1-\bt_2 \ll \bt_{1}$. Then, on expanding the heat flux in Eq.~\eqref{total rotating torque} we have
\begin{align}
    h^{(1)} &= 4 \int \frac{d\w}{2\pi }\hbar \w\sum_n \left[  (\p_\w \nu_{\bt_{1}}) n\Phi_{2,n}^{(1)} + \Delta\bt (\p_{\bt_{1}} \nu_{\beta_1}) \delta \Phi_{2,n}^{(1)} \right] \nonumber \\
    &=  -\frac{\bt_1}{\Delta \bt}\,\tau^{(1)}(0) +\mathcal{O}(\Delta\beta)\, ,\label{eq:h1}
\end{align}
where $\Phi_{2,n}^{(1),\W} = \Phi_{2,n}^{(1)} + \W\delta \Phi_{2,n}^{(1)} + \m O(\W^2)$.  The second term in the first line can be omitted since it is of order $\Delta \bt$, and in the second line we have used $\p_\w \nu_{\bt_1} = \frac{\bt_1}{\w}\p_{\bt_1} \nu_{\bt_1}  = -\frac{\bt_1}{\w\Delta \bt} \left[ \nu_{\bt_2}(\w)-\nu_{\bt_1}(\w) \right]+\mathcal{O}(\Delta\beta)$. By identifying the torque at $\Omega=0$ and small $\Delta\bt$ from Eq.~\eqref{total rotating torque}, we have obtained a form of the Onsager symmetry relation~\cite{linearresponse,nr}
\begin{align}\label{eq: onsager standard}
    \left(\frac{\p H^{(1)}}{\p \W} \right)_{\W=0,\, \bt_1=\bt_2}^{(B)} = \bt_1\left(\frac{\p\tau^{(1)}}{\p \bt_1} \right)_{\W=0, \, \bt_1 = \bt_2}^{(-B)}\,.
\end{align}
Equation~\eqref{eq: onsager standard} extends the form of the Onsager-Casimir relation \cite{Casimir1945,groot,onsager_erdmann} presented in Ref.~\cite{linearresponse} to now include nonreciprocity via an external field.  Under time reversal, the external magnetic field transforms as $B \rightarrow-B$, which then sends $\eps_f \rightarrow -\eps_f$. Since the torque in Eq.~\eqref{eq:h1} is antisymmetric in $\eps_f$, this produces an additional minus sign in Eq.~\eqref{eq: onsager standard}, making it agree with Eq.~\eqref{eq:h1}. We note that Ref.~\cite{nr} misses that extra minus sign. 


For the friction coefficient we must extract the linear term in $\W$ from Eq.~\eqref{total rotating torque}. Writing $\Phi_{2,n}^{(1),\W} = \Phi_{2,n}^{(1)} + \W\delta \Phi_{2,n}^{(1)}$, we get the friction coefficient to be a sum of two terms,
\begin{align}\label{friction 1o}
    \g^{(1)}= -4\hbar \int_0^\infty \frac{d\w}{2\pi}\sum _n &\left[  (\p_\w \nu_{\bt_1}) \, n^2 \Phi_{2,n}^{(1)} \right. \nonumber\\
    &+ \left. \left[\nu_{\bt_{2}}(\omega)-\nu_{\bt_{1}}(\omega)\right] \,n  \delta \Phi_{2,n}^{(1)}\right].
\end{align}
The latter term has a factor of $ \nu_{\bt_2 }-\nu_{\bt_1}$, so we can omit it in the following because it is of order $\Delta \beta$. The first term, the equilibrium friction, is conveniently expressed in terms of the static heat flux density, which we already have expanded in cylindrical waves. This term is analogous to the parallel-plate equilibrium friction term calculated in Ref. \cite{nr}, and we again see the correspondence $n \Leftrightarrow k_y$ and $ \Phi_{2,n}^{(1)}\Leftrightarrow S_2^{(1)}(k_y)$ between the two geometries. We note that the quantity $\delta \Phi$ in the latter terms of Eqs. (\ref{eq: onsager standard}) and (\ref{friction 1o}) cannot be diagonalized in the standard cylindrical wavefunction basis, so it is more difficult to evaluate. 

For a general nonreciprocal material, the  direction of the torque is obviously dependent on the system parameters and the external field $B_\text{ex}$. When the inner cylinder is rotated along the torque, the torque decreases, giving us a positive friction coefficient $\g^{(1)}$. While the static heat transfer is always positive, when the inner cylinder is rotated along the torque, the heat transfer decreases to first order, and one can check that $h^{(1)}\leq0$.

\subsection{Bound between heat flux, torque and friction}\label{sec: cauchy schwarz bound}
In the limit of small temperature difference, $0<\Delta \bt \ll \bt$, the quantities $H^{(1)}(0)$, $\tau^{(1)}(0)$, $h^{(1)}$ and $\g^{(1)}$ can be written as
\begin{align}
    H^{(1)}(0) &= \frac{\Delta \bt}{\bt_1}\int_0^\infty \frac{d\w}{2\pi}\sum_n \, \w^2 F(\w) \,\Phi_{2,n}^{(1)}(\w) \nonumber \\  
    \tau^{(1)}(0) &= \frac{\Delta \bt}{\bt_1}\int_0^\infty \frac{d\w}{2\pi}\sum_n \, n\w F(\w) \,\Phi_{2,n}^{(1)}(\w) \nonumber \\  
    h^{(1)} &= -\int_0^\infty \frac{d\w}{2\pi}\sum_n \, n\w F(\w) \,\Phi_{2,n}^{(1)}(\w) \nonumber \\  
    \g^{(1)} &= \int_0^\infty \frac{d\w}{2\pi}\sum_n \, n^2 F(\w) \,\Phi_{2,n}^{(1)}(\w) \, ,\label{eq:4q}
\end{align}
where we have defined $F(\w) \equiv -4\hbar (\p_\bt \nu_\bt) \geq 0$. We furthermore define, recalling that $\Phi_{2,n}^{(1)}\geq 0$,
\begin{align}
a_n(\w)\equiv \w\sqrt{F(\w)\Phi_{2,n}^{(1)}(\w)} \nonumber \\ b_n(\w)\equiv n\sqrt{F(\w)\Phi_{2,n}^{(1)}(\w)}.
\end{align}
The Cauchy-Schwarz inequality in these variables then gives
\begin{align}
\nonumber\left( \int_0^\infty \frac{d\w}{2\pi }\sum_n a_n^2(\w) \right)\left( \int_0^\infty \frac{d\w}{2\pi }\sum_n b_n^2(\w) \right)\\\geq \left( \int_0^\infty \frac{d\w}{2\pi }\sum_n a_n b_n(\w) \right)^2.
\end{align}
Inserting the expressions for $a$ and $b$ and identifying with Eq.~\eqref{eq:4q} yields
\begin{align}
    H^{(1)}(0) \g^{(1)}\geq \left|h^{(1)} \right|\left|\tau^{(1)}(0) \right|\,.\label{eq:bound2}
\end{align}
By using the analogous expressions in Table~\ref{hf table 2 geometries}, we see that Eq.~\eqref{eq:bound2} is also valid for the geometry of parallel plates studied in Ref.~\cite{nr}.

\subsection{Engine efficiency}
The engine efficiency is given by $\eta = \frac{\W  \tau^{(1)}}{H^{(1)}}$. Without loss of generality, we discuss the case where $\Omega\geq0 $ and $\tau^{(1)}\geq 0$, which yields, to leading order in $\W$, 
\begin{align}
    \eta = \W \frac{\tau^{(1)}(0)-\g^{(1)}\W}{H^{(1)}(0)+h^{(1)}\W}\,,\label{eq:eff}
\end{align}
where $\tau^{(1)}(0)$ and $H^{(1)}(0)\geq 0$ are the torque and heat flux at $\W = 0$, and the coefficients $\g^{(1)}\geq 0$ and $h^{(1)}\leq 0$ are the first-order responses derived in Eqs.~\eqref{friction 1o} and \eqref{eq:h1}, respectively. As a result, $\eta$ vanishes at $\Omega=0$ and, for small $\Omega$, grows with positive slope $\frac{\tau^{(1)}(0)}{H^{(1)}(0)}$. It then has a maximum if and only if the coefficients satisfy ${\tau^{(1)}(0)}/{\g^{(1)}} \geq {H^{(1)}(0)}/{|h^{(1)}|}$, which is precisely the inequality we have proven for small $\Delta \bt$ in Sec.~\ref{sec: cauchy schwarz bound}. To see this, we compute this maximum, which is found at
\begin{align}
\W^* &= \frac{H^{(1)}(0)}{|h^{(1)}|} \left[1-\sqrt{1- u}\right],
\end{align}
with $u \equiv \frac{\tau^{(1)}(0) \,|h^{(1)}|}{H^{(1)}(0)\,\g^{(1)}}$, which, by virtue of Eq.~\eqref{eq:bound2}, obeys $0\leq u\leq 1$, rendering $\Omega^*$ real. $\Omega^*$ is thus the optimal rotation speed at which  $\eta$ in Eq.~\eqref{eq:eff} is maximal. The maximal efficiency takes the value
\begin{align}
\eta^* &= \eta_C \,\frac{1-\sqrt{1-u}}{1+\sqrt{1-u}} \leq \eta_C\,,
\end{align}
which is therefore Carnot bounded via Eq.~\eqref{eq:bound2}.

For an explicit estimate of the engine efficiency, we consider the limit of a thin inner cylinder, so that the heat flux and torque are dominated by the terms $n=0$ and $n=\pm 1$ in Eq.~\eqref{eq:handt}. We also choose a material for which $\Phi_0$ and $\Phi_1$ have a narrow peak at a resonance frequency $\w^*$ and we define the dimensionless numbers $\al \equiv \frac{|\Phi_1-\Phi_{-1}|}{\Phi_0+\Phi_{1}+\Phi_{-1}}$ and $\sg \equiv \frac{\Phi_1+\Phi_{-1}}{\Phi_0+\Phi_{1}+\Phi_{-1}}$.  Because $\Phi_n\geq0$, we have $0\leq \al^2<\sg^2<1$. For a small nonreciprocal dielectric component $\eps_f\ll \eps_d$, we have $\al \propto \eps_f$, so $\al$ serves as a dimensionless measure of nonreciprocity. 
In the limit of small temperature difference, we then find
\begin{align}
    \frac{H^{(1)}(0)}{|h^{(1)}|}  &= \frac{\w^*\Delta \bt}{\al \bt} \nonumber \\ \frac{|\tau^{(1)}(0)|}{\g^{(1)}}&=\frac{\al\w^*\Delta \bt}{\sg \bt} \, .
\end{align}
This result yields the optimal angular speed and maximal efficiency as 
\begin{align}
    \W^* &= \frac{\w^*\eta_{\textrm{C}}}{\al}\left[1-\sqrt{1-\al^2/\sg} \right] \nonumber \\
    \eta^* &= \eta_{\textrm{C}} \frac{1-\sqrt{1-\al^2/\sg}}{1+\sqrt{1-\al^2/\sg}}< \eta_{\textrm{C}}\,.
\end{align}
In particular, for $\eps_f \ll \eps_d$, we have $\Phi_n -\Phi_{-n} \propto \eps_f\propto \al$, as shown in Sec.\ \ref{sec: small eta torque cyl}.  In this limit, $\eta^* \propto \eps_f^2\eta_{\textrm{C}}$, similarly to its parallel-plate counterpart~\cite{nr}. 

Lastly, $\eta$ grows linearly in $\Omega$ for $\Omega\ll\omega^*$, and the bound between torque and heat flux implies a bound for the initial slope. For a thin cylinder, this bound becomes $\partial_\Omega\eta_\text{cyl}|_{\Omega=0}\leq \frac{1}{\omega^*}$, and for the general case we have $\partial_\Omega\eta_\text{cyl}|_{\Omega=0}\leq \frac{N}{\omega^*}$.

\section{Conclusions}\label{sec:concl}

We have developed a framework for fluctuation-induced heat transfer and torque in enclosing configurations and applied it to a contactless cylindrical heat engine driven by nonreciprocal electromagnetic fluctuations. For a rotationally symmetric inner object, we obtain a general bound relating torque to heat transfer, and show that two rotationally symmetric reciprocal objects exhibit no net torque. Specializing to concentric cylinders, we express heat transfer and torque in terms of an angular-momentum-resolved flux density $\Phi_n(\omega)$, making explicit the correspondence with the parallel-plate geometry, where the in-plane wave vector plays an analogous role. We further show, within the proximity force approximation, that the inner cylinder is stable against displacement away from the center axis if the parallel plate pressure is repulsive. We quantify the non-equilibrium conditions that can lead to such repulsion. Finally, for a slowly rotating inner cylinder we compute the fluctuation-induced friction, and show that if used as a heat engine, the efficiency remains bounded by the Carnot limit. The latter finding is based on a bound between friction, torque, and heat transfer.

The present work suggests several directions for future study. On the theoretical side, it would be interesting to extend the torque--heat bound to more general geometries and to sharpen the estimate of the effective angular-momentum cutoff that enters the bound in practical calculations. It would also be useful to go beyond the thin-cylinder and proximity-force limits, and to analyze finite-size effects, stronger nonreciprocity, and more general material models. From the perspective of applications, one could investigate whether similar nonreciprocal fluctuation-induced torques arise in other contactless geometries, and whether the stability and efficiency of the cylindrical engine can be optimized in experimentally realistic parameter regimes. 

\subsection*{Acknowledgements}
We would like to thank Daniele Gamba for useful discussions. This research project was financially supported by the state of Lower-Saxony and the Volkswagen Foundation, Hannover, Germany (M.Kr. and D.G.) and the National Science Foundation  through grants
PHY-2209582 (N.G.) and DMR-2218849 (M.Ka.). K.A. acknowledges the hospitality of Georg-August-Universit\"at G\"ottingen.

\onecolumngrid

\appendix
\section{$2 \times 2$ cylindrical reflection tensors}\label{sec reflection matrices}
We describe here the calculation of the reflection tensor $\m T_1(n,k_z)$ for the inner cylinder. The result for the outer cylinder is computed analogously. We first calculate the dispersion relation for the dielectric in Eq.\ (\ref{eq:dielectric tensor}) by solving the eigenvalue problem $\left[\n \times \n\times (\ldots)  -\w^2\bs \eps(\w)/c^2\right]\bs u_{n,k_z} = 0$. Here, $\bs u_{n,k_z}$ is a 3-component column vector in the space of polarizations $\vct L_{n,k_z}^\alpha$, $\vct M_{n,k_z}^\alpha$, and $\vct N_{n,k_z}^\alpha$, which are defined in Eq.\ (\ref{LMN functions}). We choose to express the dispersion relation in the form $k_\rho(n,k_z,\w)$ because $n$, $k_z$, and $\w$ are conserved across the material boundaries. We consider transverse polarizations labeled as $+$ and $-$, which in the limit $\eps_f \rightarrow0$ correspond to linear combinations $\vct M_{n,k_z}^\alpha \pm \vct N_{n,k_z}^\alpha$ of the $M$ and $N$ polarizations typically used in analyzes of the reciprocal case, such as~\cite{cylradiation}.  We find the dispersion relation
\begin{align}\label{cd waves disprel}
    k_{\rho, \pm}^2 (\w, k_z) =  \frac{1+\xi}{2}\left( \frac{\w^2 \eps_d}{c^2} - k_z^2 \right) - \frac{\w^2 \eps_f^2}{2c^2\eps_d} \pm \left(\left[\frac{1+\xi}{2}\left( \frac{\w^2 \eps_d}{c^2} - k_z^2 \right) - \frac{\w^2 \eps_f^2}{2c^2\eps_d}\right]^2 - \xi\left[\left( \frac{\w^2 \eps_d}{c^2} - k_z^2 \right)^2 - \frac{\w^4\eps_f^2}{c^4} \right] \right)^{1/2} \, ,
\end{align}
where $\xi \equiv \eps_p/\eps_d$ is the anisotropy parameter. We can also obtain the eigenvectors $\bs u_\pm$ corresponding to the $+$ and $-$ waves. By considering these solutions for  $\rho<R_1$ and the vacuum solution for $\rho >R_1$, we match boundary conditions to find the $2\times 2$ reflection tensor. The homogeneous solution in the absence of the inner cylinder is regular at the origin, and hence we must consider the incident and transmitted waves to be regular, while the reflected wave is outgoing.

We define the vectors
\begin{align}\label{eq:stuv}
    \vct s_\pm^{\textrm{r}} &= \left( \frac{in}{R_1}J_{n,\pm} \,\,\,\,  -k_{\rho,\pm} J_{n,\pm}' \,\,\,\, -\frac{nk_z}{k_\mu R_1}J_{n,\mu}\right)^T  \, , \, 
    \vct t_\pm^{\textrm{r}} =  \left(ik_z \, \, \, \, 0 \, \, \, \, \frac{k_{\rho,\pm}^2}{k_\pm} \right)^T \,   J_{n,\pm} \, , \nonumber \\
    \vct v_\pm^{\textrm{r}} &= \left( 0  \, \,\,\, -\frac{nk_zc}{\w R_1}J_{n,\pm}  \, \, \,\, - \frac{ck_{\rho,\pm}k_\pm}{\w}J_{n,\pm}'\right)^T
     \quad \hbox{and} \quad
    \vct w_\pm^{\textrm{r}} = \frac{ck_{\rho,\pm}^2}{\w} \left(0\,\, \, \, 0 \, \, \,\,1\right)^T \, J_{n,\pm} \,,
\end{align}
where all Bessel functions are given by $J_{n,\pm}\equiv J_n(k_{\rho,\pm}R_1)$ and $k_{\rho,\pm}$ are the radial components of the wavevector as given in Eq.\ (\ref{cd waves disprel}). The index $\textrm{r}$ indicates the regular solution obtained from the Bessel functions $J_n$. We analogously define the vectors $\vct s_\pm^{\textrm{o}}, \vct t_\pm^{\textrm{o}}, \vct v_\pm^{\textrm{o}}, \vct w_\pm^{\textrm{o}}$ for the outgoing case by making the substitution $J_n(k_{\rho,\pm}R_1)\rightarrow H_n^{(1)}(k_{\rho,\pm}R_1)$ in Eq.\ (\ref{eq:stuv}).  Next, we define two more sets of matrices
\begin{align}
    X_{n,k_z}^\al =\begin{pmatrix}
        \bs u_+ \cdot \vct s_+^\al & \bs u_- \cdot \vct s_-^\al \\
        \bs u_+ \cdot \vct t_+^\al & \bs u_- \cdot \vct t_-^\al 
    \end{pmatrix}
    \quad \hbox{and} \quad
    Y_{n,k_z}^\al =\begin{pmatrix}
        \bs u_+ \cdot \vct v_+^\al & \bs u_- \cdot \vct v_-^\al \\
        \bs u_+ \cdot \vct w_+^\al & \bs u_- \cdot \vct w_-^\al 
    \end{pmatrix} \, ,
\end{align}
and
\begin{align}\label{vacuum 2by2 matrices}
    \m J_{n,k_z} = \begin{pmatrix}
        -\sqrt{k_0^2-k_z^2} J_n' & -\frac{nk_z}{k_0 R_1}J_n \\
        0 & \frac{k_0^2-k_z^2}{k_0}J_n
    \end{pmatrix}
    \quad \hbox{and} \quad
    \m H_{n,k_z} = \begin{pmatrix}
        -\sqrt{k_0^2-k_z^2} H_n^{(1)'} & -\frac{nk_z}{k_0 R_1}H_n^{(1)} \\
        0 & \frac{k_0^2-k_z^2}{k_0}H_n^{(1)}
    \end{pmatrix} \, ,
\end{align}
where $\al \in \{\textrm{r},\textrm{o}\}$, the Bessel functions are given by $J_n\left(\sqrt{k_0^2-k_z^2}R_1\right)$, and $k_0=\w/c$ is the magnitude of the wave vector in vacuum. Using all of these matrices, we can write the reflection tensor as
\begin{align}
    \m T_1(n,k_z) = \left[\left(X_{n,k_z}^{\textrm{r}}\right)^{-1}\m H_{n,k_z} - \left(Y_{n,k_z}^{\textrm{r}}\right)^{-1}\m H_{n,k_z} \bs \sg_1 \right]^{-1} \left[  \left( X_{n,k_z}^{\textrm{r}}\right)^{-1}\m J_{n,k_z} - \left(Y_{n,k_z}^{\textrm{r}}\right)^{-1}\m J_{n,k_z} \bs \sg_1\right] \, ,\label{eq:T1}
\end{align}
where $\sg_1$ is the first Pauli matrix. Next, we consider the reflection matrix $\m T_2$ for the outer cylinder. Here, the incident and transmitted waves are outgoing, while the reflected wave is regular. Repeating the same procedure, we get
\begin{align}
    \m T_2(n,k_z) = \left[\left(X_{n,k_z}^{\textrm{o}}\right)^{-1}\m J_{n,k_z} - \left(Y_{n,k_z}^{\textrm{o}}\right)^{-1}\m J_{n,k_z} \bs \sg_1 \right]^{-1} \left[  \left( X_{n,k_z}^{\textrm{o}}\right)^{-1}\m H_{n,k_z} - \left(Y_{n,k_z}^{\textrm{o}}\right)^{-1}\m H_{n,k_z} \bs \sg_1\right] \, .
\end{align}

\section{Direct derivation of heat flux and torque as the surface integrals of the Poynting vector and stress tensor}\label{sec: hf poynting}

We first consider the heat transfer $H_1^{(2)}$ due to fluctuations in the inner cylinder that are absorbed by the outer cylinder, which can be expressed as an integral of the normal component of the Poynting vector over a surface $\Sigma$ enclosing the outer cylinder. We choose this surface to be a cylinder $\Sigma_2$ of radius $R
_2-\ep$, giving
\begin{align}\label{heat surface integral}
    H_1^{(2)} &= \int_{-\infty}^\infty \frac{d\omega}{2\pi} \int_{\Sigma_2} \hat n \cdot   \frac{c}{4\pi \w}\,\text{Im} \, \bb \bs E \times \n \times \bs E^\dagger \rr_{\w,1} \,,
\end{align}
where $\hat n$ is the surface normal. A similar result can be obtained for the torque $\tau_1^{(2)}$ on the outer cylinder, where the force on an infinitesimal volume of the cylinder $d\m V(\vct r)$ is expressed as a surface integral of the normal component of the stress tensor $T_{ij}\equiv E_i E_j+B_i B_j -\frac{1}{2}\delta_{ij}\left( E^2 + B^2\right)$ over an enclosing surface $d\Sigma (\vct r)$. As a result, the torque on this volume is $d\bs \tau(\vct r) = \int_{d\Sigma (\vct r)} \vct r \times (T\cdot \hat n) $ and the total torque is obtained by summing over the volumes $d\m V(\vct r)$. In this sum, the contributions from common surfaces separating neighboring volumes cancel out due to the sign flip in $\hat n$. Neglecting contributions from spatial infinity, we are left with a simple surface integral over the cylinder $\Sigma_2$, 
\begin{align}\label{torque surface integral}
    \tau_1^{(2)}  = \frac{R_2}{4\pi} \int_{\Sigma_2} \hat \rho \cdot\int_{-\infty}^{\infty} \frac{d\w}{2\pi}  \text{Re} \bb \bs E \otimes \bs E^\dagger + \bs B \otimes \bs B^\dagger \rr_{\w,1} \cdot \hat \phi \, .
\end{align}
This expression can also be derived directly from the continuum volume integral, $\bs \tau = \int_{\m V_2} \vct r \times \vct f = \int_{\m V_2 } \vct r \times (\n \cdot T)$, where the force density $\vct f$ is expressed as the divergence of the stress tensor. Since $T$ is symmetric, we can write $\vct r \times  (\n \cdot T) = \n \cdot (\vct r \times T)$, and use the divergence theorem to get back the surface integral in Eq.\ (\ref{torque surface integral}).

Note that both the surface integrals are independent of $\phi$ and $z$, so the integrand is a constant and can be evaluated at any point $\vct r_0$ on the surface $\Sigma_2$. Since the averages in Eqs. (\ref{heat surface integral}) and (\ref{torque surface integral}) are quadratic in the electric field, they can be expressed in terms of the EE correlator for fluctuations originating from the inner object, namely $\mb C_{\w,1}\equiv\bb \bs E \otimes \bs E^\dagger\rr_{\w,1}$, as given by Eq.\ (\ref{eq:rad}).

Next, we expand the correlator in the cylindrical wave basis of Sec.\ \ref{sec: cyl wave basis}. This is easily done using the definitions of the $\mb T$ operators in Eq.\ (\ref{eq: G0T operator to T matrix}). We get
\begin{align}\label{correlator in LMN}
    \bra{\vct r} \mb C_{\w,1} \ket{\vct r'} = &\,a_{\bt_1}(\w)  \sum_{n,P,P'}  \int\frac{dk_z}{2\pi} \left[ \m A_{n,k_z}^{P'P} P_{n,k_z}'^{o}(\vct r)\otimes P_{n,k_z}^{o*}(\vct r') + \m D_{n,k_z}^{P'P} P_{n,k_z}^{'r}(\vct r)\otimes P_{n,k_z}^{r*}(\vct r') \right. \nonumber \\
    &\left. +\left(\m B_{n,k_z}^{P'P} P_{n,k_z}'^{o*}(\vct r)\otimes P_{n,k_z}^{r}(\vct r') + \text{h.c.} \right)  \right]\,. \\
    \m A \equiv  &\, \frac{1}{1-\m T_1 \m T_2} \left[ \frac{\m T_1+\m T_1^\dagger }{2} + \Theta^\text{pr}\m T_1 \m T_1^\dagger \right]\frac{1}{1-\m T_2^\dagger \m T_1^\dagger} \, , \quad \m B \equiv \m T_2 \m A \, , \, \m D \equiv \m T_2 \m A \m T_2^\dagger \, ,
\end{align}
where $P$ and $P'$ are summed over $\{M,N\}$, the allowed polarizations in vacuum. Finally, we use the explicit expressions for the cylinder wave functions in position space, Eq.\ (\ref{LMN functions}), along with Eq.\ (\ref{correlator in LMN}), in Eqs. (\ref{heat surface integral}) and (\ref{torque surface integral}). We note the useful vacuum relations $\n \times \vct M_n^\al = \w \vct N_n^\al/c$ and $\n \times \vct N_n^\al = \w \vct M_n^\al/c$.

We demonstrate here the calculation for the heat flux. The torque calculation follows analogously. From Eq.\ (\ref{correlator in LMN}), there are four kinds of terms, coming from $\m A,$ $\m D,$ $\m B,$ and $\m B^\dagger$, each of which must be considered separately in the propagating and evanescent regimes in order to extract the relevant imaginary part as per Eq.\ (\ref{heat surface integral}). For example, the $\m A$ terms in $\int_{\Sigma_2}\hat n \cdot \bb \bs E \times \n \times \bs E^\dagger\rr_{\w,1}$ are nonzero only in the propagating regime and evaluate to
\begin{align}
      &2\pi R_2L \sum_n\int \frac{dk_z}{2\pi}\,\Theta^\text{pr} \frac{\w}{c} \,\text{Im}\left[ \text{Tr} \left[ \m A_n(k_z)\right] \hat \rho \cdot  \left( M_n^{*o}\times N_n^{\textrm{o}} + N_n^{*o}\times M_n^{\textrm{o}}\right)\right] \nonumber \\
    = \,&2\pi R_2L \sum _n \int \frac{dk_z}{2\pi} \, \Theta^\text{pr} \text{Tr}\left[\m A_n(k_z)\right] \, k_\rho \text{Im} \left[ H_n^{(1)} \left(H_n^{(2)}\right)'-\left(H_n^{(1)}\right)'H_n^{(2)}\right] 
    = -8L\sum_n \int \frac{dk_z}{2\pi} \Theta^\text{pr} \text{Tr}\left[\m A_n \right] \,.
\end{align}
Note that the cylinder functions are all evaluated at a point $\vct r_0 \in \Sigma_2$, so that the Hankel functions above have argument $k_\rho R_2$ (where, as usual, $k_\rho \equiv \sqrt{\w^2/c^2 -k_z^2}$). In the second line, we have used the Wronskian relation satisfied by the Bessel and Hankel functions~\cite{dlmf}. The $\m D$ terms evaluate to zero, and the propagating and evanescent contributions from the $\m B$ terms respectively are
\begin{align}
    \text{Propagating terms }&= 2\pi R_2L\sum_n\int\frac{dk_z}{2\pi}  k_\rho \text{Im} \left[ \left( H_nJ_n' - H_n'J_n \right) \text{Tr} (\m B) + \left( H_n^*J_n'-H_n^{'*}J_n \right)\text{Tr}(\m B^\dagger) \right] \nonumber \\
    &= -8L\sum_n\int \frac{dk_z}{2\pi}\text{Re Tr}(\m B) \, , \\
    \text{Evanescent terms }&= 2\pi R_2L \sum_n\int \frac{dk_z}{2\pi}\frac{2i|k_\rho|}{\pi} \text{Im} \left[ \left( I_nK_n'-K_nI_n' \right) \text{Tr} (\m B) + i\left( I_nK_n'-K_nI_n' \right)\text{Tr}(\m B^\dagger) \right] \nonumber \\
    &= -8L\sum_n\int \frac{dk_z}{2\pi}\text{Re Tr}(\m B) \, ,
\end{align}
where $H_n \equiv H_n^{(1)}(k_\rho R_2)$, $I_n \equiv I_n(|k_\rho|R_2)$, and $K_n \equiv K_n(|k_\rho|R_2)$ are the Hankel function of the first kind and the modified Bessel functions of the first and second kinds, respectively. In the evanescent regime, $k_\rho = i\sqrt{k_z^2-\w^2/c^2}$ is purely imaginary, and the above expressions contain its modulus $|k_\rho|$. 

We explicitly see the contour independence of the integrals in Eqs.\ (\ref{heat surface integral}) and (\ref{torque surface integral}), since the radius $R_2$ of our surface drops out from the expressions through the Wronskian relations. Putting all the terms together, we retrieve the expressions presented in Eq.\ (\ref{eq: cylinder hf torque}).  

\section{Cylindrical heat flux densities $\Phi_{i,n}^{(j)}(\w)$}\label{sec cyl hfd}
The cylindrical heat flux density is expressed in terms of the reflection matrices $\m T_j(n,k_z)$ as
\begin{align}\label{hfd 12 in T matrix}
    \Phi_{1,n}^{(2)}(\w) = L\int\frac{dk_z}{2\pi } \text{Tr} \left[ \left(\frac{\m T_2^\dagger+\m T_2}{2} + \Theta^\text{pr} \right) \frac{1}{1-\m T_1 \m T_2} \left( \frac{\m T_1 + \m T_1^\dagger}{2} + \Theta^\text{pr}\m T_1 \m T_1^\dagger\right) \frac{1}{1-\m T_2^\dagger \m T_1^\dagger}\right] \, ,
\end{align}
where $\Theta^\text{pr}\equiv \Theta(\w/c-|k_z|)$ projects onto propagating modes. The other heat flux densities $\Phi_{i,n}^{(j)}$ can all be written in similar forms. For example, 
\begin{align}\label{hfd 22 in T matrix}
    \Phi_{2,n}^{(2)}(\w) = -L\int\frac{dk_z}{2\pi } \text{Tr} \left[ \left(\frac{\m T_1^\dagger+\m T_1}{2} + \Theta^\text{pr}\m T_1^\dagger \m T_1 \right) \frac{1}{1-\m T_2 \m T_1} \left( \frac{\m T_2 + \m T_2^\dagger}{2} + \Theta^\text{pr}\right) \frac{1}{1-\m T_1^\dagger \m T_2^\dagger}\right] \, .
\end{align}

\section{Reflection tensor for the thin inner cylinder in the small nonreciprocity limit}\label{app: thin cyl reften}
We use the same procedure as in Appendix \ref{sec reflection matrices} to calculate the reflection tensor in the limit of a thin cylinder with small nonreciprocity. We set $\eps_p = \eps_d$, as mentioned in the main text. Henceforth, we let $k_\rho = \sqrt{\w^2/c^2-k_z^2}$ and $q_\rho = \sqrt{\w^2 \eps_d/c^2 - k_z^2}$ respectively be the radial wavevector in vacuum and the dielectric in the reciprocal limit. For small nonreciprocity, $\eps_f\ll \eps_d$, we first obtain the dispersion relations and eigendirections for the $+$ and $-$ regular waves in the cylinder,
\begin{align}
    k_{\rho, \pm }^2 \equiv q_\rho^2 \pm \frac{\eps_f k_z \w}{\eps_d^{1/2}c} -\frac{\w^2 \eps_f^2}{2\eps_d c^2}+\m O(\eps_f^3) 
    \quad \hbox{and} \quad
    k_{\pm} \equiv \sqrt{k_{\rho,\pm}^2+k_z^2}= \frac{\w \eps_d^{1/2}}{c} \pm \frac{\eps_f k_z}{2\eps_d}\,+\m O(\eps_f^2).
\end{align}
The eigenvectors $\bs u_{\pm}$ simplify to
\begin{align}
    \bs u_{\pm} = \begin{pmatrix}
        0 \\ 1\\ \pm 1
    \end{pmatrix} + \eps_f \frac{c^2 q_\rho^2}{\w^2 \eps ^2} \begin{pmatrix}
         -i \\ 0 \\ \frac{\w \eps_d^{1/2}}{2ck_z}
    \end{pmatrix}\, + \m O(\eps_f^2),
\end{align}
and the general solution for the electric field just inside the dielectric boundary is given by
\begin{align}
    \vct E_\text{in} &= A^+ \left[ \vct M_{n,k_z}^{\textrm{r}}(k_{\rho,+}R)+\vct N_{n,k_z}^{\textrm{r}}(k_{\rho,+}R) + \eps_f \vct {\m N}_{n,k_z}^{\textrm{r}}(q_{\rho}R)\right] \cr & + A^- \left[ \vct M_{n,k_z}^{\textrm{r}}(k_{\rho, -}R)-\vct N_{n,k_z}^{\textrm{r}}(k_{\rho,-}R) + \eps_f \vct {\m N}_{n,k_z}^{\textrm{r}}(q_{\rho}R)\right] + \m O(\eps_f^2)\nonumber ,
\end{align}
where we have defined a new vector wavefunction
\begin{align}
    \vct {\m N}_{n,k_z}^{\textrm{r}} &\equiv \frac{1}{2}\left(\frac{cq_\rho}{\w \eps_d}\right)^2 \left[ -iq_\rho J_n'(q_\rho \rho) \hat \rho + \frac{n}{\rho} J_n(q_\rho \rho )\hat \phi + \frac{\w^2\eps_d/c^2+k_z^2}{k_z^2}J_n(q_\rho \rho) \hat z\right]
\end{align}
obeying
\begin{align}
\n \times \vct {\m N} _{n,k_z}^{\textrm{r}} \equiv \frac{q_\rho^2}{2\eps_d k_z} \vct M_{n,k_z}^{\textrm{r}} \,.
\end{align}

We must expand all these terms to first order in $\eps_f$ to get the general solution for the $E$ and $H$ fields inside the dielectric. The vacuum solutions have already been discussed in Appendix \ref{sec reflection matrices}. Imposing boundary conditions on the continuity of the $\hat \phi$ and $\hat z$ components of the $E$ and $H$ fields, we get eight linear equations for the unknown reflection and transmission coefficients $\m T^{PP'}$ and $A^{\pm P}$ with $P,P'\in \{M,N\}$, which are written in Eq.\ (\ref{eq: 8 linear eqs}) of the main text. Each of these reflection coefficients can be expanded in powers of $\eps_f$ as $\m T^{PP'}_{n,k_z}=\sum_{r=0}^\infty  \m T_{n,k_z,(r)}^{PP'}\eps_f^r$ and the matrix $\m K_n(\eps_f)$ can be expanded as $\m K_n(\eps_f) =  K_{n,(0)} + \eps_f  K_{n,(1)} + \m O(\eps_f^2)$, where
\begin{align}\label{K0 K1 expressions}
    \m K_{n,(0)} &= \begin{pmatrix}
        x_n & y_n & p_n +p'_n & p_n -p'_n \\
        0 & z_n & r_n & -r_n \\
        y_n & x_n & \eps_d^{1/2}(p_n+p'_n) & -\eps_d^{1/2}(p_n-p'_n) \\
        z_n & 0 & \eps_d^{1/2} r_n & \eps_d^{1/2} r_n
    \end{pmatrix}  
     \quad \hbox{and} \quad
    \m K_{n,(1)} = \frac{k_z}{2\eps}\begin{pmatrix}
        0 & 0 & s_n+s'_n & s_n-s'_n \\
        0 & 0 & t_n & t_n \\
        0 & 0 & \eps_d^{1/2}(u_n +u'_n) & \eps_d^{1/2}(u_n - u'_n) \\
        0 & 0 & \eps_d^{1/2}v_n & -\eps_d^{1/2}v_n
    \end{pmatrix} \,,
\end{align}
with
\begin{align}\label{eq: xyzpqrstuv defs}
    x_n &\equiv  -k_\rho H_n'(k_\rho R)  \,, \, 
    y_n \equiv -\frac{nk_zc}{R\w}H_n(k_\rho R) \,, \, 
    z_n \equiv \frac{ck_\rho^2}{\w}H_n(k_\rho R) , \, \nonumber \\
    p_n &\equiv -q_\rho J_n'(q_\rho R) \,, \, 
    p'_n \equiv -\frac{nk_zc}{R\w \eps_d^{1/2}}J_n(q_\rho R)\, , \, 
    r_n \equiv \frac{cq_\rho^2}{\w \eps_d^{1/2}} J_n (q_\rho R) 
    \,,  \\
    s_n &\equiv \frac{n}{k_zR}J_n(q_\rho R) - \frac{nk_z}{q_\rho} J_n'(q_\rho R), \, 
    s_n' \equiv  -\frac{\w\eps_d^{1/2}}{c} \frac{n^2 - q_\rho^2 R^2}{q_\rho^2 R} J_n(q_\rho R) \, , \, 
    t_n \equiv q_\rho R J_n'(q_\rho R) +\frac{\w^2 \eps_d/c^2 + k_z^2}{k_z^2}J_n(q_\rho R)\,, \nonumber \\
    u_n &\equiv -\frac{\w \eps_d^{1/2}}{cq_\rho} \left[ \frac{n^2 - q_\rho^2 R^2}{q_\rho R} J_n(q_\rho R) + \frac{q_\rho^2}{k_z^2} J_n'(q_\rho R) \right]\,,\, 
    u_n' \equiv -\frac{nk_z}{q_\rho} J_n'(q_\rho R)\, , 
    \quad \hbox{and} \quad
    v_n \equiv q_\rho R J_n'(q_\rho R)+2J_n(q_\rho R)\,.
\end{align}

To take the thin-cylinder limit, $k_\rho R \ll 1$ and $q_\rho R \ll 1$, we can use the asymptotic expressions $J_0(x)\simeq 1$ and $H_0(x)\simeq \frac{2i}{\pi}\ln x$ for $n=0$
and $J_n (x) \simeq \frac{1}{n!}\left(\frac{x}{2} \right)^n$ and $H_n(x) \simeq \frac{(n-1)!}{i\pi} \left(\frac{2}{x} \right)^n$ for $n \geq 1$~\cite{dlmf}. It is easy to see that the contribution to $\m T^{PP'}$ from the $n^\text{th}$ mode $\varphi_n$ is less than $\m O(R^{-n-1})$, so that the main contributions to the heat transfer and torque come from the $n = 0, \pm 1$ modes. We take the asymptotic limits in Eqs. (\ref{K0 K1 expressions}) and (\ref{eq: xyzpqrstuv defs}) and explicitly calculate the elements of the reflection matrix using Eq.\ (\ref{eq: 8 linear eqs}). For $n=0$ we get Eq.\ (\ref{eq: n=0 oef2}) in the main text. The $n=0$ modes only contribute to the heat transfer, and  the dominating term is from the TE component $\m T^{NN}$~\cite{nr, thesisgolyk}. As expected from symmetries, these terms are only affected at quadratic order in nonreciprocity.

For $n = \pm 1$, we can use Eq.\ (\ref{eq: n=0,1 K symmetries}) to write the reflection matrix elements as
\begin{align}\label{eq: T decompos}
    \m T_{\pm 1, k_z}^{PP} &= \m T_{1,(0)}^{PP} \pm \m T_{1,(1)}^{PP}  \quad \hbox{and} \quad
    \m T_{\pm1, k_z}^{P \overline P} = \pm \m T_{1,(0)}^{P\overline P}+\m T_{1,(1)}^{P\overline P}\, , \, P \in\{M,N\},  
\end{align}
where we define $\overline P$ such that $\overline P=N$ if $ P=M$ and $\overline P=M$ if $P=N$. It thus suffices to consider $n=+1$. We first invert the matrix $\m K_{(0)}$ to show that
\begin{align}\label{eq: T matrix 0 order}
    \m T_{1,(0)}^{MM}=\frac{i\pi \w^2 R^2}{4c^2}\frac{\eps_d-1}{\eps_d+1}\,,\, 
    \m T_{1,(0)}^{NN} = \frac{i\pi k_z^2 R^2}{4 }\frac{\eps_d-1}{\eps_d+1}\,,
     \quad \hbox{and} \quad
    \m T_{1,(0)}^{MN} = -\m T_{1,(0)}^{NM} = \frac{i\pi \w k_z}{4c} \frac{\eps_d-1}{\eps_d+1}\, .
\end{align}
The reciprocal transmission coefficients for the $+$ and $-$ waves are then
\begin{align}
    A_{(0)}^{\pm M} = \frac{k_\rho}{2q_\rho } \frac{(1+\eps_d)k_z \pm 2\eps_d^{1/2}\w/c}{(1+\eps_d)\left(k_z \pm \eps_d^{1/2}\w/c \right)}
    \quad \hbox{and} \quad
    A_{(0)}^{\pm N} = \frac{k_\rho}{2q_\rho } \frac{\pm (1+\eps_d)\w/c +2\eps_d^{1/2}k_z}{(1+\eps_d)\left( \eps_d^{1/2}\w/c \pm k_z\right)}\,.
\end{align}
As discussed in the main text, the correction to the $\m T$-matrix at first order in nonreciprocity is obtained from the matrix $-\eps_f\m K_{(0)}^{-1}\m K_{(1)}\m K_{(0)}^{-1}$. Using this result, one can relate the first-order reflection coefficients $\m T_{1,(1)}^{PP'}$, defined in Eq.\ (\ref{eq: T decompos}), to the zeroth order transmission coefficients $A_{(0)}^{\pm P}$. After a tedious but straightforward calculation, we obtain
\begin{align}\label{eq: nonrecip T matrix 1 order}
    \begin{pmatrix}
        \m T_{1,(1)}^{MM} & \m T_{1,(1)}^{MN} \\
        \m T_{1,(1)}^{NM} & \m T_{1,(1)}^{NN}
    \end{pmatrix} = 
    -\frac{i\pi R^2\eps_f}{4 (1+\eps_d)^2} \begin{pmatrix}
        \w^2/c^2 & \w k_z /c \\
        \w k_z /c & k_z^2
    \end{pmatrix}\,.
\end{align}
Putting together Eqs. (\ref{eq: T decompos}), (\ref{eq: T matrix 0 order}), and (\ref{eq: nonrecip T matrix 1 order}), we arrive at the result in the main text. 

\section{Drude-Lorentz model for nonreciprocal plates}
\label{app:DL}
The Drude-Lorentz model with coupling to an external magnetic field $B\hat{z}$ gives the following elements of $ \bs \eps_i $ in Eq.\ (\ref{eq:dielectric tensor}), ~\cite{Milton2023}
\begin{subequations}
\begin{align}
& \epsilon_{p,i} = 1 + \frac{\omega_{\textrm{p}i}^2}{\omega_{0i}^2-\omega^2-i\gamma_i\omega},\label{eps_p}\\
& \epsilon_{d,i} = 1 + \frac{\omega_{\textrm{p}i}^2\left(\omega_{0i}^2-\omega^2-i\gamma_i\omega\right)}{\left(\omega_{0i}^2-\omega^2-i\gamma_i\omega\right)^2-\left(\omega_{\textrm{c}}\omega\right)^2}\label{eps_d}\\
& \eps_{f,i} = -\frac{\omega_{\textrm{p}i}^2\omega_c\omega}{\left(\omega_{0i}^2-\omega^2-i\gamma_i\omega\right)^2-\left(\omega_{\textrm{c}}\omega\right)^2},\label{eta}
\end{align}
\end{subequations}
where $\omega_{\textrm{p}i}$, $\omega_{0i}$, and $\gamma_i$ are the plasma frequency, resonance frequency of $\epsilon_{p,i}$, and damping rate respectively, and $ \omega_{\textrm{c}} \propto B $ is the cyclotron frequency, which determines the strength of the nonreciprocity. We consider identical plasma frequencies and damping rates for both plates ($ \omega_{\textrm{p}i} \equiv \omega_{\textrm{p}} $  and $ \gamma_i \equiv \gamma $).

\begin{figure}[!htbp]
\begin{center}
\includegraphics[width=0.48\linewidth]{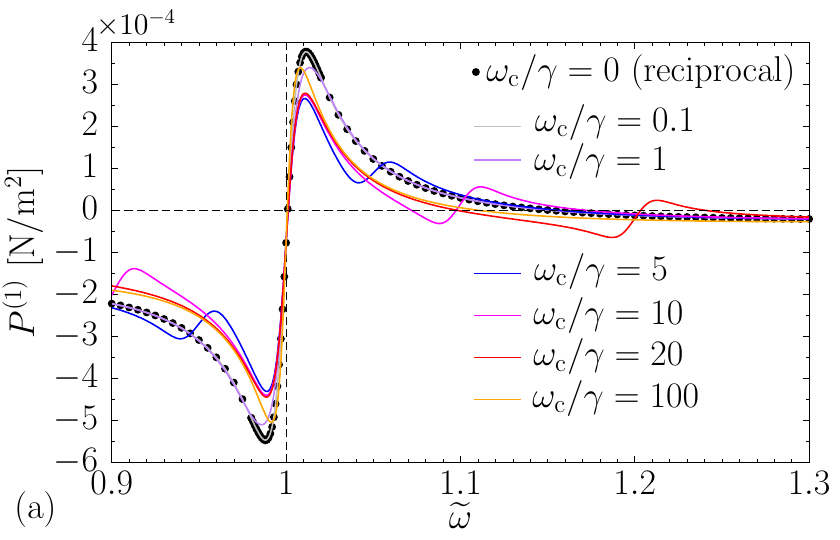}
\includegraphics[width=0.48\linewidth]{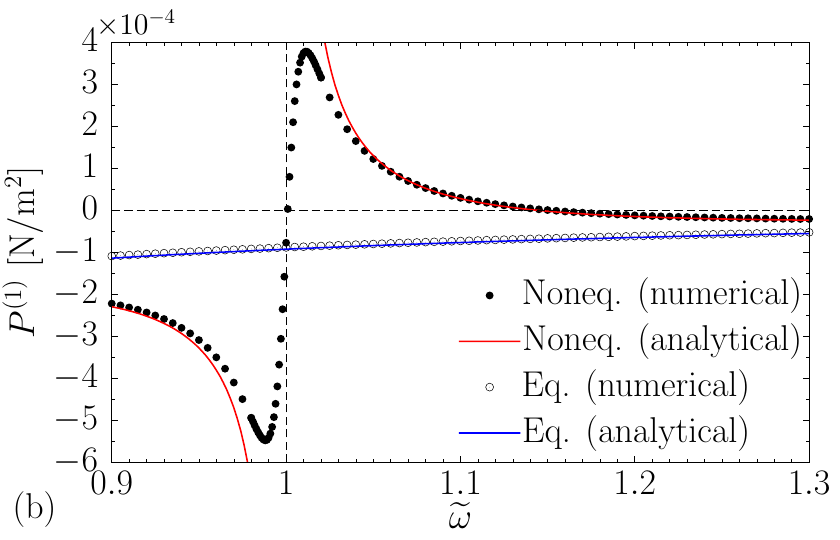}
\end{center}
\caption{\label{fig:Pressure_plates}\textbf{(a):} Total pressure acting on plate 1 (at temperature $ T_1 = 400 \ \textrm{K} $) in the presence of plate 2 ($ T_2 = 300 \ \textrm{K} $) as a function of the ratio of the bare resonant frequencies, for different values of the cyclotron frequency $ \omega_{\textrm{c}} $, which determines the strength of nonreciprocity (external magnetic field). The separation distance is $ d = 100 \ \textrm{nm} $ and we have chosen $ \omega_{\textrm{p}} = 2 \times 10^{12} \ \textrm{rad} \ \textrm{s}^{-1} $, $ \gamma = 10^{11} \ \textrm{rad} \ \textrm{s}^{-1} $, and $ \omega_2 = 10^{13} \ \textrm{rad} \ \textrm{s}^{-1} $. A positive (negative) pressure corresponds to repulsion from (attraction to) plate 2. Dashed lines are included as guides to the eye. \textbf{(b):} Comparison between the pressure (for $ \omega_{\textrm{c}} = 0 $) and its analytical approximation given by Eq.~\eqref{eq:P1an}. The equilibrium pressure is plotted for $ T_1 = T_2 = 300 \ \textrm{K} $.}
\end{figure}

Figure~\ref{fig:Pressure_plates}(a) illustrates the nonequilibrium pressure on plate 1, computed numerically from Eqs.~\eqref{eq:P1separation}--\eqref{eq:Pi1}, as a function of the ratio $ \widetilde{\omega} \equiv \omega_1/\omega_2 $ of the bare ($\omega_{\textrm{c}} = 0$) resonance frequencies $\omega_i = \sqrt{\omega_{0i}^2+\omega_{\textrm{p}}^2/2}$. The pressure can be repulsive (positive) if $ \widetilde{\omega} > 1 $ for $ \widetilde{\beta} \equiv T_1/T_2 > 1 $. While small $ \omega_{\textrm{c}} $ has little effect, the behavior changes significantly when $ \omega_{\textrm{c}} $ is comparable to the other frequencies: The repulsive pressure is lowered overall, but may appear at larger $ \widetilde{\omega} $ (where it is already attractive for reciprocal plates) due to an additional maximum. The equilibrium pressure was found to remain attractive under the influence of nonreciprocity, but its amplitude decreases as $ \omega_{\textrm{c}} $ increases.

To gain an analytical understanding for small $B$, we expand the Fresnel coefficient in Eq.~\eqref{eq:rNN},
\begin{equation}
r_i^{NN}(\omega,\theta) = -\frac{\frac{1}{2}\omega_{\textrm{p}}^2}{\omega^2-\omega_i^2+i\gamma\omega} + \omega_{\textrm{c}}g_i(\omega)\sin\theta + \omega_{\textrm{c}}^2\left(p_i(\omega)+q_i(\omega)\sin^2\theta\right) + \mathcal{O}\left(\omega_{\textrm{c}}^3\right),
\label{eq:rNN_DL}
\end{equation}
whose imaginary and real parts are
\begin{subequations}
\begin{align}
& \textrm{Im}\left[r_i^{NN}\right] = \frac{\frac{\omega_{\textrm{p}}^2}{2}\gamma\omega}{\left(\omega^2-\omega_i^2\right)^2+\gamma^2\omega^2} + \omega_{\textrm{c}}\textrm{Im}\left[g_i(\omega)\right]\sin\theta + \omega_{\textrm{c}}^2\left(\textrm{Im}\left[p_i(\omega)\right]+\textrm{Im}\left[q_i(\omega)\right]\sin^2\theta\right) + \mathcal{O}\left(\omega_{\textrm{c}}^3\right),\label{eq:rNN_DL_Im}\\
& \textrm{Re}\left[r_i^{NN}\right] = -\frac{\frac{\omega_{\textrm{p}}^2}{2}\left(\omega^2-\omega_i^2\right)}{\left(\omega^2-\omega_i^2\right)^2+\gamma^2\omega^2} + \omega_{\textrm{c}}\textrm{Re}\left[g_i(\omega)\right]\sin\theta + \omega_{\textrm{c}}^2\left(\textrm{Re}\left[p_i(\omega)\right]+\textrm{Re}\left[q_i(\omega)\right]\sin^2\theta\right) + \mathcal{O}\left(\omega_{\textrm{c}}^3\right).\label{eq:rNN_DL_Re}
\end{align}
\end{subequations}
Using $ \textrm{Li}_3\left(r_1^{NN}r_2^{NN}\right) \approx r_1^{NN}r_2^{NN} $, Eqs.~\eqref{eq:P0} and~\eqref{eq:Pi1} involve integration of $ \textrm{Im}\left[r_i^{NN}\right]\textrm{Re}\left[r_{\overline{i}}^{NN}\right] $. In the product of the reciprocal terms in Eqs.~\eqref{eq:rNN_DL_Im} and~\eqref{eq:rNN_DL_Re},
if the peak in $ \textrm{Im}\left[r_i^{NN}\right] $ is sharp ($ \gamma \ll \omega_i $) and $ \textrm{Re}\left[r_{\overline{i}}^{NN}\right] $ changes only slightly within this peak ($ |\omega_i-\omega_{\overline{i}}| \gg \gamma$), it can be approximated as a Lorentzian, which we may further approximate by a small window function $\widetilde\delta$ with unit integral, \cite{nr} 
\begin{equation}
\frac{\frac{\omega_{\textrm{p}}^2}{2}\gamma\omega}{\left(\omega^2-\omega_i^2\right)^2+\gamma^2\omega^2} \approx \frac{\pi\omega_{\textrm{p}}^2}{4\omega_i}\widetilde{\delta}(\omega-\omega_i)
\label{eq:DeltaApproximation}.
\end{equation}
The integral over $\omega$ is then straightforwardly performed as with a conventional delta distribution: $\widetilde{{\delta}}(\omega-\omega_i)$ selects $ \textrm{Re}\left[r_{\overline{i}}^{NN}\right] $ at $\omega_i$. This and consideration of nonreciprocal terms in Eqs.~\eqref{eq:rNN_DL_Im} and~\eqref{eq:rNN_DL_Re} lead to the formula in Eq.~\eqref{eq:P1an}. As demonstrated in Fig.~\ref{fig:Pressure_plates}(b), this formula is in good agreement with the numerical results. Remarkably, in equilibrium, the formula works even if $\widetilde{\omega}$ is arbitrarily close or equal to unity.


\twocolumngrid
\bibliography{apssamp}

@article{Casimir1945,
  author  = {Casimir, H. B. G.},
  title   = {On Onsager's Principle of Microscopic Reversibility},
  journal = {Reviews of Modern Physics},
  volume  = {17},
  pages   = {343--350},
  year    = {1945},
  doi     = {10.1103/RevModPhys.17.343}
}

@misc{shah2026,
      title={A Contactless Heat Engine Driven by Nonreciprocal Fluctuation-Induced Torques}, 
      author={Dhruv Shah and Kiryl Asheichyk and David Gelbwaser-Klimovsky and Noah Graham and Mehran Kardar and Matthias Krüger},
      year={2026},
      eprint={2606.25053},
      archivePrefix={arXiv},
      primaryClass={quant-ph},
      url={https://arxiv.org/abs/2606.25053}, 
}

@book{Tsang_Mie,
        author    = {L. Tsang and J. A. Kong and K. H. Ding},
        title     = {Scattering of Electromagnetic Waves, Theories and Applications},
        publisher = {Wiley},
        year      = 2000
        }

@article{Strekha22,
  title = {Trace expressions and associated limits for nonequilibrium Casimir torque},
  author = {Strekha, Benjamin and Molesky, Sean and Chao, Pengning and Kr\"uger, Matthias and Rodriguez, Alejandro W.},
  journal = {Phys. Rev. A},
  volume = {106},
  issue = {4},
  pages = {042222},
  numpages = {19},
  year = {2022},
  month = {Oct},
  publisher = {American Physical Society},
  doi = {10.1103/PhysRevA.106.042222},
  url = {https://link.aps.org/doi/10.1103/PhysRevA.106.042222}
}

@book{Jackson,
    author = {Jackson, John David},
    title = {Classical electrodynamics},
    edition = {3rd Edition},
    publisher = {Wiley},
    year = {1998},
    isbn = {978-0-471-30932-1},
}

@article{HT_exp2,
	author = {C.M Hargreaves},
	doi = {10.1016/0375-9601(69)90264-3},
	issn = {0375-9601},
	journal = {Phys. Lett. A},
	number = {9},
	pages = {491},
	title = {Anomalous radiative transfer between closely-spaced bodies},
	url = {https://www.sciencedirect.com/science/article/pii/0375960169902643},
	volume = {30},
	year = {1969}
}

@article{HT_exp3,
  title = {Near-Field Heat Transfer in a Scanning Thermal Microscope},
  author = {Kittel, Achim and M\"uller-Hirsch, Wolfgang and Parisi, J\"urgen and Biehs, Svend-Age and Reddig, Daniel and Holthaus, Martin},
  journal = {Phys. Rev. Lett.},
  volume = {95},
  issue = {22},
  pages = {224301},
  numpages = {4},
  year = {2005},
  month = {Nov},
  publisher = {American Physical Society},
  doi = {10.1103/PhysRevLett.95.224301},
  url = {https://link.aps.org/doi/10.1103/PhysRevLett.95.224301}
}

@article{HT_exp4,
author = {Yusuke Kajihara and Keishi Kosaka and Susumu Komiyama},
journal = {Opt. Express},
number = {8},
pages = {7695--7704},
publisher = {Optica Publishing Group},
title = {Thermally excited near-field radiation and far-field interference},
volume = {19},
month = {Apr},
year = {2011},
url = {https://opg.optica.org/oe/abstract.cfm?URI=oe-19-8-7695},
doi = {10.1364/OE.19.007695},
}

@article{shorttrace,
  title = {Nonequilibrium Electromagnetic Fluctuations: Heat Transfer and Interactions},
  author = {Kr\"uger, Matthias and Emig, Thorsten and Kardar, Mehran},
  journal = {Phys. Rev. Lett.},
  volume = {106},
  issue = {21},
  pages = {210404},
  numpages = {4},
  year = {2011},
  month = {May},
  publisher = {American Physical Society},
  doi = {10.1103/PhysRevLett.106.210404},
  url = {https://link.aps.org/doi/10.1103/PhysRevLett.106.210404}
}

@article{nr,
  title = {Near Field Propulsion Forces from Nonreciprocal Media},
  author = {Gelbwaser-Klimovsky, David and Graham, Noah and Kardar, Mehran and Kr\"uger, Matthias},
  journal = {Phys. Rev. Lett.},
  volume = {126},
  issue = {17},
  pages = {170401},
  numpages = {5},
  year = {2021},
  month = {Apr},
  publisher = {American Physical Society},
  doi = {10.1103/PhysRevLett.126.170401},
  url = {https://link.aps.org/doi/10.1103/PhysRevLett.126.170401}
}

@article{gyrodparticle,
author = {Guo, Yu and Fan, Shanhui},
title = {Single Gyrotropic Particle as a Heat Engine},
journal = {ACS Photonics},
volume = {8},
number = {6},
pages = {1623-1629},
year = {2021},
doi = {10.1021/acsphotonics.0c01920},
url = {https://doi.org/10.1021/acsphotonics.0c01920},
OPTeprint = {https://doi.org/10.1021/acsphotonics.0c01920}
}

@misc{reid2017,
      title={Photon Torpedoes and {R}ytov Pinwheels: Integral-Equation Modeling of Non-Equilibrium Fluctuation-Induced Forces and Torques on Nanoparticles}, 
      author={M. T. Homer Reid and O. D. Miller and A. G. Polimeridis and A. W. Rodriguez and E. M. Tomlinson and S. G. Johnson},
      eprint={1708.01985},
      archivePrefix={arXiv},
      year={2017}
}

@article{Muller_2016,
  title = {Anisotropic particles near surfaces: Propulsion force and friction},
  author = {M\"uller, Boris and Kr\"uger, Matthias},
  journal = {Phys. Rev. A},
  volume = {93},
  issue = {3},
  pages = {032511},
  numpages = {12},
  year = {2016},
  month = {Mar},
  publisher = {American Physical Society},
  doi = {10.1103/PhysRevA.93.032511},
  url = {https://link.aps.org/doi/10.1103/PhysRevA.93.032511}
}

@article{Kruger_2011,
    doi = {10.1209/0295-5075/95/21002},
    url = {https://dx.doi.org/10.1209/0295-5075/95/21002},
    year = {2011},
    month = {jun},
    publisher = {},
    volume = {95},
    number = {2},
    pages = {21002},
    author = {M. Krüger and T. Emig and G. Bimonte and M. Kardar},
    title = {Non-equilibrium {C}asimir forces: Spheres and sphere-plate},
    journal = {Europhys. Lett.}
}

@article{Polder_HT_plates,
  title = {Theory of Radiative Heat Transfer between Closely Spaced Bodies},
  author = {Polder, D. and Van Hove, M.},
  journal = {Phys. Rev. B},
  volume = {4},
  issue = {10},
  pages = {3303--3314},
  numpages = {0},
  year = {1971},
  month = {Nov},
  publisher = {American Physical Society},
  doi = {10.1103/PhysRevB.4.3303},
  url = {https://link.aps.org/doi/10.1103/PhysRevB.4.3303}
}

@article{trace,
  title = {Trace formulas for nonequilibrium Casimir interactions, heat radiation, and heat transfer for arbitrary objects},
  author = {Kr\"uger, Matthias and Bimonte, Giuseppe and Emig, Thorsten and Kardar, Mehran},
  journal = {Phys. Rev. B},
  volume = {86},
  issue = {11},
  pages = {115423},
  numpages = {26},
  year = {2012},
  month = {Sep},
  publisher = {American Physical Society},
  doi = {10.1103/PhysRevB.86.115423},
  url = {https://link.aps.org/doi/10.1103/PhysRevB.86.115423}
}

@techreport{rytov1959theory,
  title={Theory of electric fluctuations and thermal radiation},
  author={Rytov, Sergei Mikhailovich},
  institution={Defense Technical Information Center},
  year={1959}
}

@article{rahi,
  title = {Scattering theory approach to electrodynamic Casimir forces},
  author = {Rahi, Sahand Jamal and Emig, Thorsten and Graham, Noah and Jaffe, Robert L. and Kardar, Mehran},
  journal = {Phys. Rev. D},
  volume = {80},
  issue = {8},
  pages = {085021},
  numpages = {27},
  year = {2009},
  month = {Oct},
  publisher = {American Physical Society},
  doi = {10.1103/PhysRevD.80.085021},
  url = {https://link.aps.org/doi/10.1103/PhysRevD.80.085021}
}

@misc{dlmf,
        title = "{\it NIST Digital Library of Mathematical Functions}",
howpublished = "\url{https://dlmf.nist.gov/}, Release 1.2.4 of 2025-03-15",
        year = "2025",
        url = "https://dlmf.nist.gov/",
        note = "F.~W.~J. Olver, A.~B. {Olde Daalhuis}, D.~W. Lozier, B.~I. Schneider,
                R.~F. Boisvert, C.~W. Clark, B.~R. Miller, B.~V. Saunders,
                H.~S. Cohl, and M.~A. McClain, eds."}

@article{fanfan,
  title = {Nonreciprocal radiative heat transfer between two planar bodies},
  author = {Fan, Lingling and Guo, Yu and Papadakis, Georgia T. and Zhao, Bo and Zhao, Zhexin and Buddhiraju, Siddharth and Orenstein, Meir and Fan, Shanhui},
  journal = {Phys. Rev. B},
  volume = {101},
  issue = {8},
  pages = {085407},
  numpages = {9},
  year = {2020},
  month = {Feb},
  publisher = {American Physical Society},
  doi = {10.1103/PhysRevB.101.085407},
  url = {https://link.aps.org/doi/10.1103/PhysRevB.101.085407}
}

@article{cylradiation,
  title = {Heat radiation from long cylindrical objects},
  author = {Golyk, Vladyslav A. and Kr\"uger, Matthias and Kardar, Mehran},
  journal = {Phys. Rev. E},
  volume = {85},
  issue = {4},
  pages = {046603},
  numpages = {15},
  year = {2012},
  month = {Apr},
  publisher = {American Physical Society},
  doi = {10.1103/PhysRevE.85.046603},
  url = {https://link.aps.org/doi/10.1103/PhysRevE.85.046603}
}

@article{linearresponse,
  title = {Linear response relations in fluctuational electrodynamics},
  author = {Golyk, Vladyslav A. and Kr\"uger, Matthias and Kardar, Mehran},
  journal = {Phys. Rev. B},
  volume = {88},
  issue = {15},
  pages = {155117},
  numpages = {6},
  year = {2013},
  month = {Oct},
  publisher = {American Physical Society},
  doi = {10.1103/PhysRevB.88.155117},
  url = {https://link.aps.org/doi/10.1103/PhysRevB.88.155117}
}

@misc{nearfieldSI,
    author = "Gelbwaser-Klimovsky, David and Graham, Noah and Kardar, Mehran and Kruger, Matthias",
    title = "Supplementary material for Near Field Propulsion Forces from Nonreciprocal Media",
    year="2021"
}

@mastersthesis{movingcorr,
    author = "Rauch, Philip",
    title = "Dynamical Casimir interactions and their connection to nonreciprocal media ",
    school = "Institute for Theoretical Physics
Georg-August-Universitat, 37077 Gottingen",
    year = "2022"
}

@phdthesis{thesisgolyk,
    author = "Golyk, Vladyslav A.",
    title = "Non-Equilibrium Fluctuation Induced-Phenomena in Quantum Electrodynamics",
    school = "Massachusetts Institute of Technology",
    year = "2014"
}

@article{greenkubo,
doi = {10.1088/0034-4885/29/1/306},
url = {https://doi.org/10.1088/0034-4885/29/1/306},
year = {1966},
month = {jan},
publisher = {},
volume = {29},
number = {1},
pages = {255},
author = {R Kubo},
title = {The fluctuation-dissipation theorem},
journal = {Reports on Progress in Physics},
}

@ARTICLE{cas,
   author       = "H. B. G. Casimir ",
   year         = "1948",
   journal      = "Proc. K. Ned. Akad. Wet.",
   volume       = "B 51",
   pages        = "793"
}

@ARTICLE{liftz,
   author       = "E. M. Lifshitz", 
   year         = "1956", 
   journal      = "Zh. Eksp. Teor. Fiz. ", 
   volume       = "29", 
   pages        = "94"
}

@article{biggerheat,
author = {Shen, Sheng and Narayanaswamy, Arvind and Chen, Gang},
title = {Surface Phonon Polaritons Mediated Energy Transfer between Nanoscale Gaps},
journal = {Nano Letters},
volume = {9},
number = {8},
pages = {2909-2913},
year = {2009},
doi = {10.1021/nl901208v},
    note ={PMID: 19719110}
}

@article{objectinsideobject,
  title = {Scattering theory approach to electrodynamic Casimir forces},
  author = {Rahi, Sahand Jamal and Emig, Thorsten and Graham, Noah and Jaffe, Robert L. and Kardar, Mehran},
  journal = {Phys. Rev. D},
  volume = {80},
  issue = {8},
  pages = {085021},
  numpages = {27},
  year = {2009},
  month = {Oct},
  publisher = {American Physical Society},
  doi = {10.1103/PhysRevD.80.085021},
  url = {https://link.aps.org/doi/10.1103/PhysRevD.80.085021}
}

@INBOOK{wengchochew,
  author={Chew, Weng Cho},
  booktitle={Waves and Fields in Inhomogenous Media}, 
  title={Cylindrically and Spherically Layered Media}, 
  year={1995},
  publisher={IEEE Press},
  volume={},
  number={},
  pages={161-209},
  doi={10.1109/9780470547052.ch3}
}

@article{eckhardt,
  title = {Macroscopic theory of electromagnetic fluctuations and stationary radiative heat transfer},
  author = {Eckhardt, W.},
  journal = {Phys. Rev. A},
  volume = {29},
  issue = {4},
  pages = {1991--2003},
  numpages = {0},
  year = {1984},
  month = {Apr},
  publisher = {American Physical Society},
  doi = {10.1103/PhysRevA.29.1991},
  url = {https://link.aps.org/doi/10.1103/PhysRevA.29.1991}
}

@article{zhu_persistent_2016,
    title = {Persistent directional current at equilibrium in nonreciprocal many-body near field electromagnetic heat transfer},
    volume = {117},
    number = {13},
    journal = {Physical Review Letters},
    author = {Zhu, Linxiao and Fan, Shanhui},
    year = {2016},
    pages = {134303},
}

@article{Henkes2025,
  title = {Propulsion force and heat transfer for nonreciprocal nanoparticles},
  author = {Henkes, Laila and Asheichyk, Kiryl and Kr\"uger, Matthias},
  journal = {Phys. Rev. B},
  volume = {111},
  issue = {3},
  pages = {035441},
  numpages = {13},
  year = {2025},
  month = {Jan},
  publisher = {American Physical Society},
  doi = {10.1103/PhysRevB.111.035441},
  url = {https://link.aps.org/doi/10.1103/PhysRevB.111.035441}
}

@article{zhu_near-complete_2014,
    title = {Near-complete violation of detailed balance in thermal radiation},
    volume = {90},
    number = {22},
    journal = {Physical Review B},
    author = {Zhu, Linxiao and Fan, Shanhui},
    year = {2014},
    pages = {220301},
}

@article{herz_green-kubo_2019,
    title = {Green-{Kubo} relation for thermal radiation in non-reciprocal systems},
    volume = {127},
    number = {4},
    journal = {EPL},
    author = {Herz, F. and Biehs, S. A.},
    year = {2019},
    pages = {44001},
}

@article{ben-abdallah_photon_2016,
    title = {Photon thermal hall effect},
    volume = {116},
    number = {8},
    journal = {Physical Review Letters},
    author = {Ben-Abdallah, Philippe},
    year = {2016},
    pages = {084301},
}

@article{latella_giant_2017,
    title = {Giant thermal magnetoresistance in plasmonic structures},
    volume = {118},
    number = {17},
    journal = {Physical Review Letters},
    author = {Latella, Ivan and Ben-Abdallah, Philippe},
    year = {2017},
    pages = {173902},
}

@article{caloz_electromagnetic_2018,
    title = {Electromagnetic nonreciprocity},
    volume = {10},
    number = {4},
    journal = {Physical Review Applied},
    author = {Caloz, Christophe and Alù, Andrea and Tretyakov, Sergei and Sounas, Dimitrios and Achouri, Karim and Deck-Léger, Zoé-Lise},
    year = {2018},
    pages = {047001},
}

@article{asadchy_tutorial_2020,
    title = {Tutorial on {Electromagnetic} {Nonreciprocity} and {Its} {Origins}},
    volume = {108},
    number = {10},
    journal = {Proceedings of the IEEE},
    author = {Asadchy, Viktar and Mirmoosa, Mohammad S. and Diaz-Rubio, Ana and Fan, Shanhui and Tretyakov, Sergei A.},
    year = {2020},
    pages = {1684},
}

@book{Bordag2009,
    author = {Bordag, Michael and Klimchitskaya, Galina Leonidovna and Mohideen, Umar and Mostepanenko, Vladimir Mikhaylovich},
    title = {{Advances in the Casimir Effect}},
    publisher = {Oxford University Press, New York},
    year = {2009},
    isbn = {9780199238743},
    doi = {10.1093/acprof:oso/9780199238743.001.0001},
    url = {https://doi.org/10.1093/acprof:oso/9780199238743.001.0001},
}

@article{Zhao2009,
  title = {Repulsive {C}asimir Force in Chiral Metamaterials},
  author = {Zhao, R. and Zhou, J. and Koschny, Th. and Economou, E. N. and Soukoulis, C. M.},
  journal = {Phys. Rev. Lett.},
  volume = {103},
  issue = {10},
  pages = {103602},
  numpages = {4},
  year = {2009},
  month = {Sep},
  publisher = {American Physical Society},
  doi = {10.1103/PhysRevLett.103.103602},
  url = {https://link.aps.org/doi/10.1103/PhysRevLett.103.103602}
}

@article{Grushin2011,
  title = {Tunable {C}asimir Repulsion with Three-Dimensional Topological Insulators},
  author = {Grushin, Adolfo G. and Cortijo, Alberto},
  journal = {Phys. Rev. Lett.},
  volume = {106},
  issue = {2},
  pages = {020403},
  numpages = {4},
  year = {2011},
  month = {Jan},
  publisher = {American Physical Society},
  doi = {10.1103/PhysRevLett.106.020403},
  url = {https://link.aps.org/doi/10.1103/PhysRevLett.106.020403}
}

@article{Castillo-Lopez2022,
  title = {Casimir forces out of thermal equilibrium near a superconducting transition},
  author = {Castillo-L\'opez, S. G. and Esquivel-Sirvent, R. and Pirruccio, G. and Villarreal, C.},
  journal = {Sci. Rep.},
  volume = {12},
  pages = {2905},
  year = {2022},
  doi = {10.1038/s41598-022-06866-5},
  url = {https://doi.org/10.1038/s41598-022-06866-5}
}

@article{Bimonte2011,
  title = {Dilution and resonance-enhanced repulsion in nonequilibrium fluctuation forces},
  author = {Bimonte, Giuseppe and Emig, Thorsten and Kr\"uger, Matthias and Kardar, Mehran},
  journal = {Phys. Rev. A},
  volume = {84},
  issue = {4},
  pages = {042503},
  numpages = {4},
  year = {2011},
  month = {Oct},
  publisher = {American Physical Society},
  doi = {10.1103/PhysRevA.84.042503},
  url = {https://link.aps.org/doi/10.1103/PhysRevA.84.042503}
}

@article{Iizuka2023,
  title = {Implications of reciprocity for the spectra of equilibrium and nonequilibrium {C}asimir forces},
  author = {Iizuka, Hideo and Fan, Shanhui},
  journal = {Phys. Rev. B},
  volume = {108},
  issue = {7},
  pages = {075429},
  numpages = {12},
  year = {2023},
  month = {Aug},
  publisher = {American Physical Society},
  doi = {10.1103/PhysRevB.108.075429},
  url = {https://link.aps.org/doi/10.1103/PhysRevB.108.075429}
}

@article{Antezza2008,
  title = {Casimir-Lifshitz force out of thermal equilibrium},
  author = {Antezza, Mauro and Pitaevskii, Lev P. and Stringari, Sandro and Svetovoy, Vitaly B.},
  journal = {Phys. Rev. A},
  volume = {77},
  issue = {2},
  pages = {022901},
  numpages = {22},
  year = {2008},
  month = {Feb},
  publisher = {American Physical Society},
  doi = {10.1103/PhysRevA.77.022901},
  url = {https://link.aps.org/doi/10.1103/PhysRevA.77.022901}
}

@article{Milton2023,
  title = {Vacuum torque, propulsive forces, and anomalous tangential forces: Effects of nonreciprocal media out of thermal equilibrium},
  author = {Milton, Kimball A. and Guo, Xin and Kennedy, Gerard and Pourtolami, Nima and DelCol, Dylan M.},
  journal = {Phys. Rev. A},
  volume = {108},
  issue = {2},
  pages = {022809},
  numpages = {15},
  year = {2023},
  month = {Aug},
  publisher = {American Physical Society},
  doi = {10.1103/PhysRevA.108.022809},
  url = {https://link.aps.org/doi/10.1103/PhysRevA.108.022809}
}

@article{Rahi2010,
  title = {Constraints on Stable Equilibria with Fluctuation-Induced (Casimir) Forces},
  author = {Rahi, Sahand Jamal and Kardar, Mehran and Emig, Thorsten},
  journal = {Phys. Rev. Lett.},
  volume = {105},
  issue = {7},
  pages = {070404},
  numpages = {4},
  year = {2010},
  month = {Aug},
  publisher = {American Physical Society},
  doi = {10.1103/PhysRevLett.105.070404},
  url = {https://link.aps.org/doi/10.1103/PhysRevLett.105.070404}
}

@book{Feynman1963,
        author    = {R. P. Feynman and R. B. Leighton and M. Sands},
        title     = {The Feynman Lectures on Physics},
        publisher = {Addison-Wesley, Boston},
        year      = {1963},
        volume = {III}
        }

@article{PhysRevLett.97.160401,
  title = {Opposites Attract: A Theorem about the Casimir Force},
  author = {Kenneth, Oded and Klich, Israel},
  journal = {Phys. Rev. Lett.},
  volume = {97},
  issue = {16},
  pages = {160401},
  numpages = {4},
  year = {2006},
  month = {Oct},
  publisher = {American Physical Society},
  doi = {10.1103/PhysRevLett.97.160401},
  url = {https://link.aps.org/doi/10.1103/PhysRevLett.97.160401}
}

@mastersthesis{onsager_erdmann,
    author = "Erdmann, Samuel",
    title = "Onsager symmetries of non-reciprocal media",
    school = "Institute for Theoretical Physics
Georg-August-Universitat, 37077 Gottingen",
    year = "2021",
    type = "Bachelor's Thesis"
}

@book{groot,
  author    = {de Groot, S. R. and Mazur, P.},
  title     = {Non-Equilibrium Thermodynamics, Chapter VII},
  publisher = {North-Holland Publishing Company},
  address   = {Amsterdam},
  year      = {1962}
}

@misc{Jiao2026,
      author={Liyao Jiao and Yaohua Liu and Gaomin Tang},
      eprint={2606.22746},
      year={2026},
      archivePrefix={arXiv}
}

\end{document}